\documentclass[fleqn,onecolumn,11pt]{wlscirep}
\usepackage{bm}
\usepackage{color}
\usepackage{color}
\usepackage{alltt,dsfont}
\usepackage{appendix}
\usepackage{amsmath,amssymb}
\usepackage{braket}
\usepackage[capitalise]{cleveref}
\usepackage{ulem}
\usepackage{float}
\usepackage{graphicx,subfigure,bbm}

\usepackage{lipsum}
\usepackage{nameref}
\usepackage{hyperref}
\usepackage{cleveref}

\usepackage[export]{adjustbox}
\usepackage{subscript}
\usepackage{titlesec,setspace}

\newcommand{\onlinecite}[1]{\hspace{-1 ex} \nocite{#1}\citenum{#1}}

\newcommand{\refdisp}[1]{Ref. [\onlinecite{#1}]}
\newcommand{\figdisp}[1]{Fig. \ref{#1}}
\newcommand{\disp}[1]{Eq. (\ref{#1})}

\newcommand{\beq}{\begin{eqnarray}}
\newcommand{\eeq}{\end{eqnarray}}

\title{1/4 is the new 1/2 when topology is intertwined with Mottness}

\author[1]{Peizhi Mai}
\author[1]{Jinchao Zhao}
\author[2,3,4]{Benjamin E. Feldman}
\author[1,*]{Philip W. Phillips}

\affil[1]{Department of Physics and Institute of Condensed Matter Theory, University of Illinois at Urbana-Champaign, Urbana, IL 61801, USA\looseness=-1}
\affil[2]{Geballe Laboratory of Advanced Materials, Stanford, CA 94305, USA\looseness=-1}
\affil[3]{Department of Physics, Stanford University, Stanford, CA 94305, USA\looseness=-1}
\affil[4]{Stanford Institute for Materials and Energy Sciences, SLAC National Accelerator Laboratory, Menlo Park, CA 94025, USA\looseness=-1}

\begin{abstract}
In non-interacting systems, bands from non-trivial topology emerge strictly at half-filling and exhibit either the quantum anomalous Hall or spin Hall effects.  Here we show using determinantal quantum Monte Carlo and an exactly solvable strongly interacting model that these topological states now shift to quarter filling.  A topological Mott insulator is the underlying cause.  The peak in the spin susceptibility is consistent with a possible ferromagnetic state at $T=0$.  The onset of such magnetism would convert the quantum spin Hall to a quantum anomalous Hall effect. While such a symmetry-broken phase typically is accompanied by a gap, we find that the interaction strength must exceed a critical value for this to occur.  Hence, we predict that topology can obtain in a gapless phase but only in the presence of interactions in dispersive bands. These results explain the recent quarter-filled quantum anomalous Hall effects seen in moir\'e systems. 

\end{abstract}
\date{\today}

\begin{document}

\flushbottom
\maketitle

\section*{Introduction}

Although topological insulators\cite{kanemele1,kanemele2,bhz,bz2006,rahulroy2,jmoore,Fu1,Fu2,Qi1,Qi2,Konig1,Hsieh1,Hsieh2,Xia,Chen,HasanKane} represent a new class of bulk insulating materials with gapless conducting edges, their physics is completely entailed by the band theory of non-interacting electrons.  The new twist is that should two atoms reside in each unit cell, the standard insulating gap that obtains at half-filling, full lower band, does not tell the whole story when spin-orbit coupling\cite{kanemele1,kanemele2,bhz} is present. As long as time-reversal invariance is maintained, two spinful counter-propagating edge modes exist and exhibit a quantized conductance proportional to $e^2/h$, thereby giving rise to a quantum spin Hall (QSH) effect in two dimensions. Within the Kane-Mele (KM)\cite{kanemele1,kanemele2} and Bernevig-Hughes-Zhang (BHZ)\cite{bhz} models, the QSH effect obtains only at half-filling. In a general non-interacting system, this physics obtains at a filling equal to the inverse number of atoms per unit cell, $1/q$.   This physics is robust to perturbations that yield only  smooth deformations\cite{HasanKane} of the Hamiltonian.  Additionally, the quantum anomalous Hall (QAH) effect, that is, the existence of a quantized Hall conductance with zero net magnetic field, also requires half-filling of the Haldane model\cite{Haldane}.  
As the QAH effect breaks time-reversal symmetry while the QSH effect does not, it is difficult for them to be realized in the same material.

However, recently, both effects\cite{WenjinZhao,TingxinLi} have been observed in the same material in direct contrast to predictions of standard non-interacting models\cite{kanemele1,kanemele2,bhz}.  In the AB-moir\'e-stacked transition metal dichalcogenide (TMD) bilayer MoTe$_2$/WSe$_2$\cite{WenjinZhao,TingxinLi}, the QSH insulator is observed at $\nu=2$ with the QAH effect residing at $\nu=1$. To date, this constitutes the first observation of the intertwining of these effects in the same material and hence the question of the minimal model required to explain the conflation of both is open.  In terms of the 4-band KM/BHZ model, $\nu=2$ and $\nu=1$ correspond to half-filling and quarter-filling, respectively.  Numerous theories\cite{SuYing,DongZhihuan,XieYingMing2,ChangYaoWen,XieMing,PanHaining,Devakul2,YangZhang,Devakul,YingMingXie,FengchengWu,Rademaker} have been put forth in this context, and the most recent experiment\cite{Taozui} shows that both valleys contribute to the QAH effect and hence valley coherence rather than valley polarization is the operative mechanism.  The striking deviation from the standard theory raises the question: can interactions drive either of these transitions away from half- to quarter-filling in the KM/BHZ models?  

It is this question that we address here.  We show quite generally that at a temperature above any ordering tendency, strong interactions shift the QSH effect to quarter filling with a decrease of the spin Chern number by a factor of two. However, the spin susceptibility exhibits a peak indicating a tendency to ferromagnetism as the temperature is lowered. 
 Such an ordered ground state would be consistent with the Lieb-Schultz-Mattis\cite{lsm1,lsm2} (LSM) theorem and recent exact diagonalization\cite{Yang} on one of the models treated here.  Whether or not such a ground state is gapped depends on the flatness of the band and interaction strength.  In the flat-band limit, the ferromagnetic ground state is always gapped whereas for a dispersive band, the interactions must exceed a critical value for a gap to obtain. These results raise the possibility of a gapless topological semi-metallic state with non-trivial temperature corrections to the Hall conductance\cite{Cooper1}.  Generally, We argue that when the interactions dominate, the QSH must give way to a ferromagnetic QAH state at $T=0$ at $1/4$-filling.    Since this is a generic conclusion on the most general models proven to undergird the QSH effect, we analyze the experiments\cite{TingxinLi,Taozui} in this context.  Our model yields a quarter-filled QAH effect which coexists with a QSH effect at half-filling as is seen experimentally, in the presence of a flat lower band and intermediate interaction. 


 A brief survey of interacting topological systems is in order as our key result hinges on the interplay between the two.  Most studies on the KM-Hubbard\cite{DongZheng,ShunLiYu,Hohenadler2012,DHLee} and the BHZ-Hubbard\cite{Amaricci,Yoshida2013,Tada,Budich} models focused on the half-filled system and found a transition from a QSH insulator to a topologically trivial anti-ferromagnetic Mott insulator as the interaction strength $U$ increases.  In addition, for models more relevant to flat-band twisted bilayer graphene, \refdisp{Bultinck2,Lian} have provided a strong-coupling analysis and a density-matrix-renormalization group study\cite{Soejima} has found that the gapless state at half-filling in the spinless (and hence Mottless) Bisritzer-MacDonald (BM) model\cite{BM} yields a quantum anomalous Hall state in the presence of Coulomb interactions. In an extensive\cite{Xie} exact diagonalization study on an 8-band BM model, $U(4)$ ferromagnets were observed always with the onset of a gap.  Quantum Monte Carlo\cite{Liao,Hofmann} on the spinful model reveals a series of insulating states at half-filling. In the mean-field context, models focused on layered graphene systems have addressed the origin of quantum Hall ferromagnetism in the interacting BM model\cite{YaHui,Bultinck,Bultinck2} while others have argued that a topological Mott insulators (TMI) emerges at half-filling in the presence of on-site and nearest neighbor interactions in the tight-binding model (with only nearest-neighbor hopping) on a honeycomb lattice\cite{Raghu}.  However, the latter proposal has not been substantiated by subsequent numerical studies\cite{Motruk,Varney1,Varney2,Assaad2} that have found half-filling to be a trivial Mott insulator when interactions are sufficiently large. Interactions also lie at the heart of fractional topological insulators\cite{Neupert2,bz2006,Levin,Repellin,Qi2} built from fractional Chern insulators\cite{Regnault,Tang,Neupert1,KaiSun,Sheng} which resemble the fractional quantum Hall effect but with no net magnetic field.  Such phases appear at a fractional filling in a flat-band $\Delta_0 \gg W_0$ (where $\Delta_0$ is the non-interacting topological gap and $W_0$ is the bandwidth) and require nearest-neighbor interactions. 
 A recent study on the strongly interacting spinful Haldane model\cite{mfp} demonstrates that a Chern Mott insulator originates at quarter-filling with Chern number $C=\pm1$. This physics arises as a general consequence of an interplay between Mottness and topology.  

Motivated by \refdisp{mfp,TingxinLi,Taozui}, we explore the general phenomena that emerge from the interplay between Mottness and the QSH effect in the context of the KM and BHZ models.  To demonstrate that the quarter-filled state is a TMI with a strongly correlated QSH effect, we numerically solve both the KM-Hubbard and BHZ-Hubbard Hamiltonians using determinantal quantum Monte Carlo (DQMC) as well as dynamical cluster approximation (DCA) and construct an analytically solvable Hamiltonian for a general interacting QSH system and obtain consistent results for sufficiently large interactions. 

\section*{Results}

\subsection*{Hubbard interaction}

\begin{figure*}[ht]
    \centering
    \includegraphics[width=0.8\textwidth]{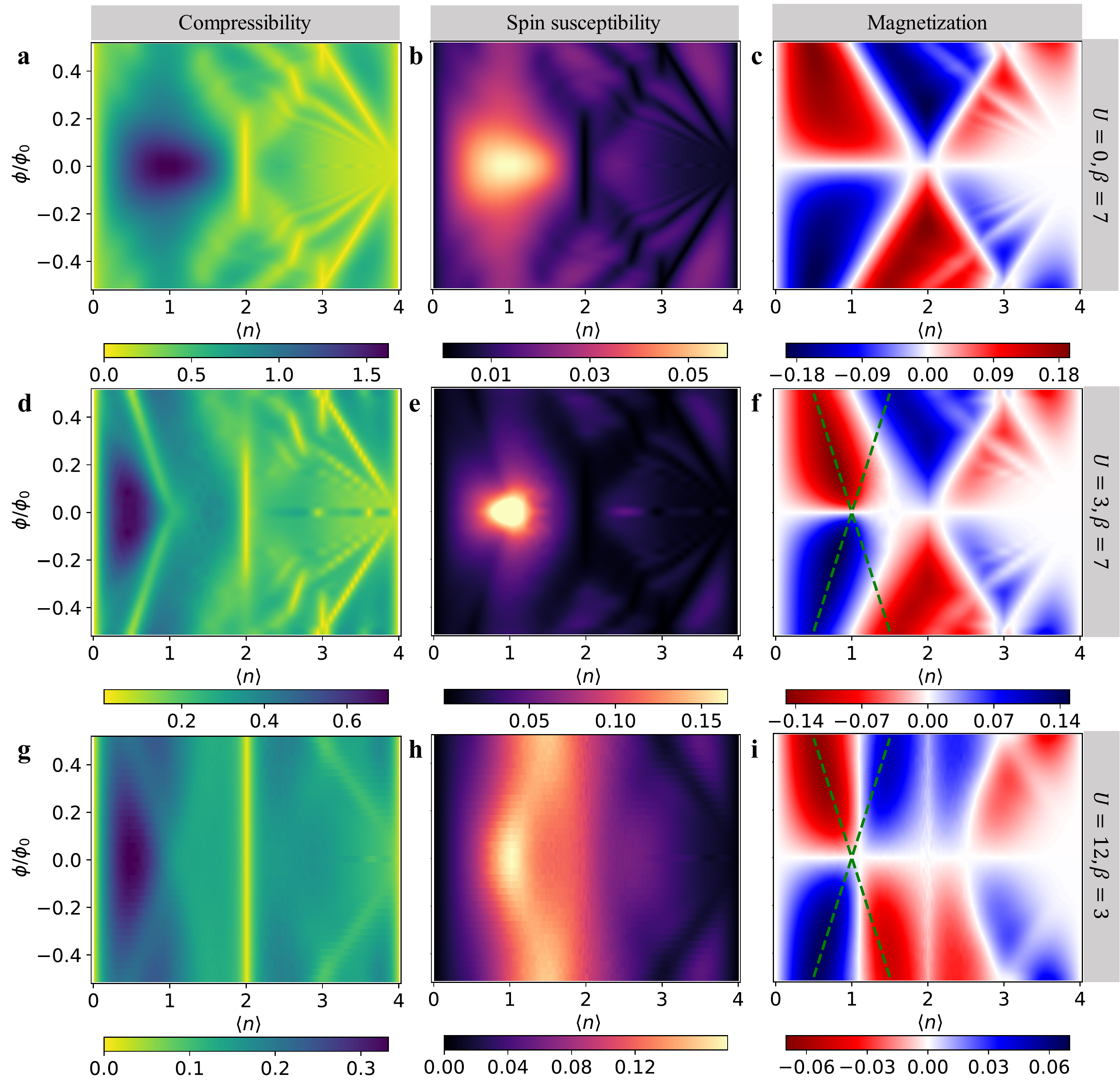}
    \caption{{\bf Compressibility, spin susceptibility and magnetization of the flat-band generalized KM-HH model}. DQMC results for the flat-band generalized KM-HH model ($t^\prime=0.3, \psi=0.81$) at $U=0, \beta=7/t$ ({\bf a}-{\bf c}), $U=3t, \beta=7/t$ ({\bf d}-{\bf f}) and $U=12t, \beta=3/t$ ({\bf g}-{\bf i}). In each row, the compressibility, spin susceptibility and magnetization are presented in order from left to right as a function of magnetic flux and electron density. The dashed green lines in panels {\bf f} and {\bf i} serve as a guide to the eye for the crossing pairs of Landau levels in the QSH effect.
    }
    \label{fig:KMHflat}
\end{figure*}

The DQMC simulation results for the generalized KM-Hofstadter-Hubbard (KM-HH) model (see Methods) on a honeycomb lattice at $\psi=0.81$ and $t^\prime/t=0.3$ are shown in \figdisp{fig:KMHflat}. For this choice of parameters, the non-interacting lower band is rather flat with bandwidth $W_{0-}\approx 0.28$ and the topological gap is $\Delta_0\approx1.62$, the upper bandwidth is $W_{0+}\approx 4.37$, where the subscript $0$ indicates non-interacting. This mimics the flat-bands in moir\'e TMD experiments. The tunability of bandwidths in the KM model (unlike the bands in the BHZ model which are always dispersive $W_{0+}=W_{0-}\geq\Delta_0$)  makes the KM model ideal for studying both flat-band and dispersive physics.  

A key quantity that helps discern the topology in the presence of a probe magnetic field is the charge compressibility,
\beq
\chi=\beta\chi_c=\frac{\beta}{N}\sum_{{\bf i},{\bf j}}\left[ \langle n_{\bf i} n_{\bf j}\rangle - \langle n_{\bf i}\rangle \langle n_{\bf j}\rangle \right], \label{charge}
\eeq
where the sublattice and spin summations are implied in $n_{\bf i}$.  Regardless of density, the inverse slope of the leading straight-line incompressible valley that extends to the zero-field limit \cite{mfp} provides the Chern number.  As a probe, this field  does not alter our claim of a QSH phase at zero field. In the non-interacting case (\figdisp{fig:KMHflat}a) at $\beta=7$, there is a short middle vertical straight line at low fields which indicates a Chern number $C_0=0$ at $\langle n\rangle=2$.  This state bifurcates into two lines or equivalently two Landau levels (LLs) at higher magnetic flux. This crossing pair of zero-mode LLs is a reliable fingerprint for the QSH effects observed in experiments\cite{Konig1}. Note the asymmetry around $\langle n\rangle=2$ arises entirely because the lower band is flat while the upper band is dispersive.  In this regime, the lines with finite slopes all represent the standard integer quantum Hall states.

The second quantity we calculate is the spin susceptibility defined as 
\beq
\chi_s=\sum_{\bf r}S({\bf r})-Nm_z^2=\frac{1}{N}\sum_{{\bf i},{\bf r}}[\langle S^z_{\bf i} S^z_{{\bf i}+{\bf r}}\rangle -\langle S^z_{\bf i}\rangle \langle S^z_{{\bf i}+{\bf r}}\rangle],\label{spin}
\eeq
where $m_z=\sum_{\bf i}\langle S^z_{\bf i}\rangle/N$ is the magnetization per spin. The non-interacting spin susceptibility is related to the compressibility by $\chi_s=\chi/(4\beta)$ as shown in \figdisp{fig:KMHflat}(b) with reverse color scale. \figdisp{fig:KMHflat}(c) shows the magnetization. Even though the Zeeman field is absent, a non-zero Peierls flux can magnetize the system since the spin-up and -down electron bands have different Chern numbers. The non-interacting results at lower temperatures can be found in the supplement. What we alert the reader to is the absence of any topologically non-trivial states at $\langle n\rangle =1$. 




In the presence of interactions $U=3t$ (already strongly correlated for the lower band), the new feature and hence prediction is the emergence of a topologically non-trivial state at $\langle n\rangle =1$. In \figdisp{fig:KMHflat}d, the inverse slope of the trace extending to $\langle n\rangle=1$ is $\pm 1$ and thus gives the Chern number. The absence of the right-moving counterpart signifies a QAH effect rather than a QSH effect.   At $\langle n \rangle=2$, the standard QSH effect remains. Consequently, we have a system in which both the QAH and QSH effects obtain simply by changing the filling. For $\langle n\rangle >2$, the physics is weakly interacting as $U<W_{0+}$. The bright peak in the spin susceptibility in \figdisp{fig:KMHflat}e indicates a possible tendency for ferromagnetism at $\langle n\rangle =1$.  This is supported by the asymmetry in the dotted lines that cross at zero field and $\langle n\rangle=1$ in the magnetization in \figdisp{fig:KMHflat}f.  Such asymmetry signifies that an infinitesimal field would lead to a polarization of the spins and hence ferromagnetism.  

We then further increase the interaction strength but have to raise the temperature to $\beta=3$ due to the Fermion sign problem in DQMC (see supplement). In the final row of \figdisp{fig:KMHflat} for the compressibility when $U=12t$, which far exceeds $W_{0-}+W_{0+}+\Delta_0\approx 6$, the non-interacting QSH Landau fan vanishes for $\langle n\rangle=2$ turning into a trivial Mott insulator and most strikingly, a new LL emerges corresponding to the mirror image of the QAH state that terminates at $\langle n\rangle=1$.  The presence of both Landau components completes the high-temperature QSH features at quarter filling.  The magnetization (\figdisp{fig:KMHflat}(i)) shows a more dramatic change than does the compressibility; namely it vanishes at $\langle n\rangle=2$ as a result of the anti-ferromagnetic Mott insulator.  Further, the magnetization splits into peaks on either side of $\langle n\rangle =1$ that continues to be asymmetrical and hence is consistent with a tendency for spontaneous Ising ferromagnetism despite the presence of both LLs.  This physics in \figdisp{fig:KMHflat}(d-f) is only present in the flat-band limit when $U$ is much larger than the bandwidth but comparable to the topological gap.  Consequently, our theoretical work here is consistent with the sudden onset of the QAH state. Since the temperature for \figdisp{fig:KMHflat}(g-i) is higher than the previous row, their features are softer.  

\begin{figure}[ht]
    \centering
    \includegraphics[width=0.7\textwidth]{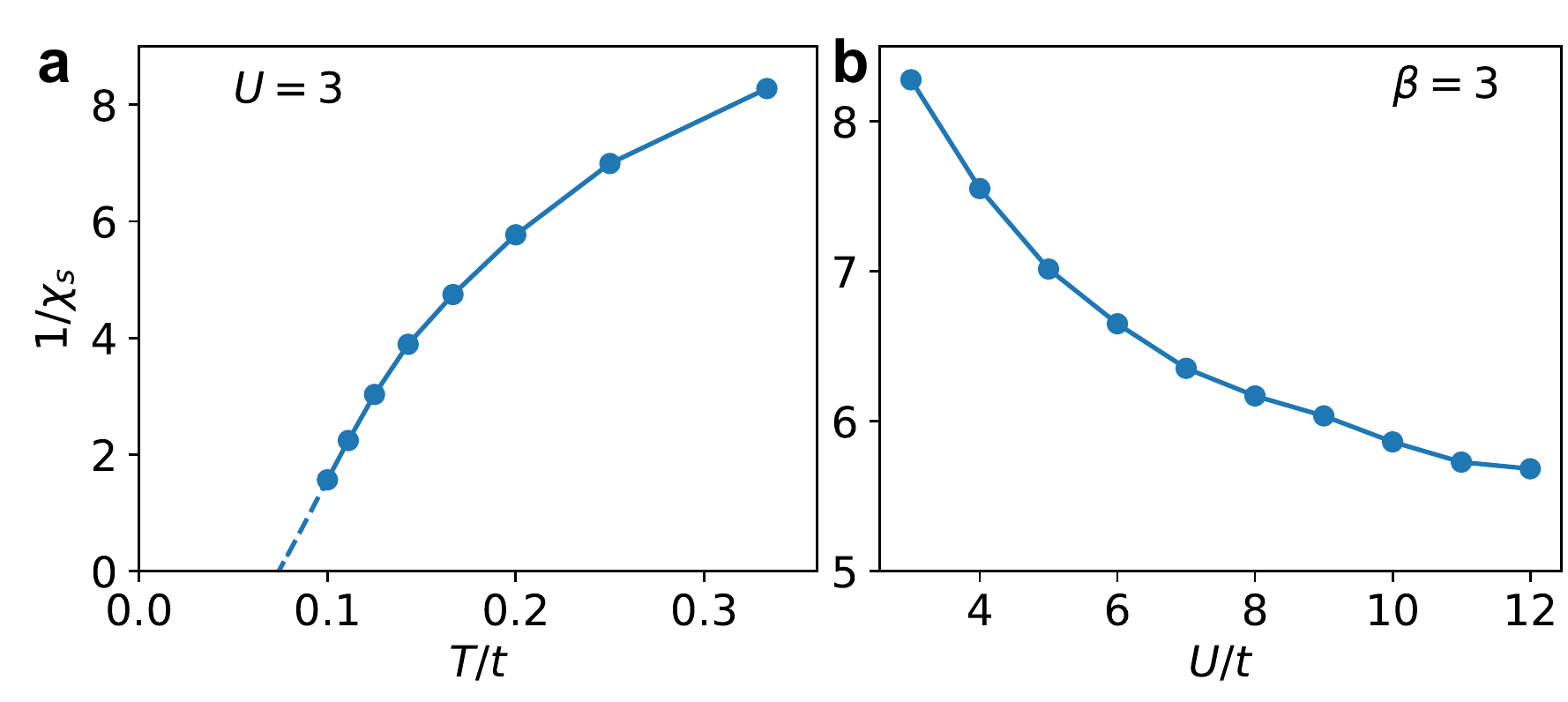}
    \caption{ {\bf Temperature evolution and $U$-dependence of the spin susceptibility}. Inverse spin susceptibility $1/\chi_s$ at quarter-filling ($\langle n \rangle=1$) of the interacting flat-band generalized KM-HH model. Panel {\bf a} contains the temperature evolution of $1/\chi_s$ at $U/t=3$ with extrapolation to zero. Panel {\bf b} shows  $1/\chi_s$ as a function of interaction strength at a fixed inverse temperature $\beta=3/t$. 
    }
    \label{fig:chis}
\end{figure}

To confirm the tendency for ferromagnetism, it is important to compute the temperature dependence of the spin susceptibility.  Shown in \figdisp{fig:chis}a is the inverse spin susceptibility as the temperature is lowered with zero external magnetic flux. Displayed clearly is a possible divergence of the susceptibility ($1/\chi_s\rightarrow0$) consistent with ordering. With extrapolation, we find that it supports a finite-temperature transition to ferromagnetism. Note that this does not violate the Mermin-Wagner theorem which forbids the spontaneous breaking of continuous symmetries at finite temperature in low-dimensional ($d\leq2$) systems with short-range interactions. In the KM-Hubbard model with spin-orbit coupling, the system no longer has the full SU(2) symmetry but only conserves $\hat{S}^z$. Then it is the Ising symmetry that is spontaneously broken in this transition and thus allowed at a finite temperature. As this is an interaction-driven effect, we expect an enhancement of the susceptibility as $U$ increases.  This is also borne out in \figdisp{fig:chis}b.  Together these figures justify our claim of interaction-driven ferromagnetism as the temperature is lowered. A ferromagnetic QAH state will stabilize at zero temperature even though QSH features could be present at high temperatures when $U$ is sufficiently large (\figdisp{fig:KMHflat}(g-i)). We also observe a similar high-temperature phenomenon in the dispersive case $\psi=0.5$ (see supplement).

To show the generality of the 1/4-filled topological state, we consider the BHZ-Hofstadter-Hubbard (BHZ-HH) model (see Methods) on a square lattice.  Note in this model, both bands are dispersive and have the same bandwidth. Without loss of generality, we set  $M/t=1$, then $W_{0-}=W_{0+}=\Delta_{0}=2t$ ($t=1$ as the energy scale). The non-interacting $1/2$-filled system is a QSH insulator with $C_s=2$. It is the spin Chern number that describes a QSH insulator. To measure this quantity, we use a spin-dependent time-reversal-invariant (TRI) magnetic field inspired by cold-atom experiments\cite{Goldman,Aidelsburger}, namely $\phi_{{\bf i},{\bf j}}\rightarrow  \sigma \phi_{{\bf i},{\bf j}}$. The compressibility measured in this way we refer to as TRI compressibility. The minus sign coupled to spin-down electrons changes the corresponding Chern number $C_{\downarrow}^{\text{TRI}}=-C_{\downarrow}$. Thus, the ``TRI" Chern number measured in the TRI compressibility  $C^{\text{TRI}}=C_{\uparrow}^{\text{TRI}}+C_{\downarrow}^{\text{TRI}}=C_{\uparrow}-C_{\downarrow}=C_s$ corresponds to the spin Chern number in the BHZ-HH model. This method overcomes the breakdown of the simple additivity formula for $C_s$ when the spin channels are mixed and ${\bf k}$ is no longer a good quantum number in the presence of interactions. Through this quantity, we can read the spin Chern number from the inverse slope of the TRI compressibility (see the supplement for the non-interacting examples).

\begin{figure*}[ht!]
    \centering
    \includegraphics[width=1\textwidth]{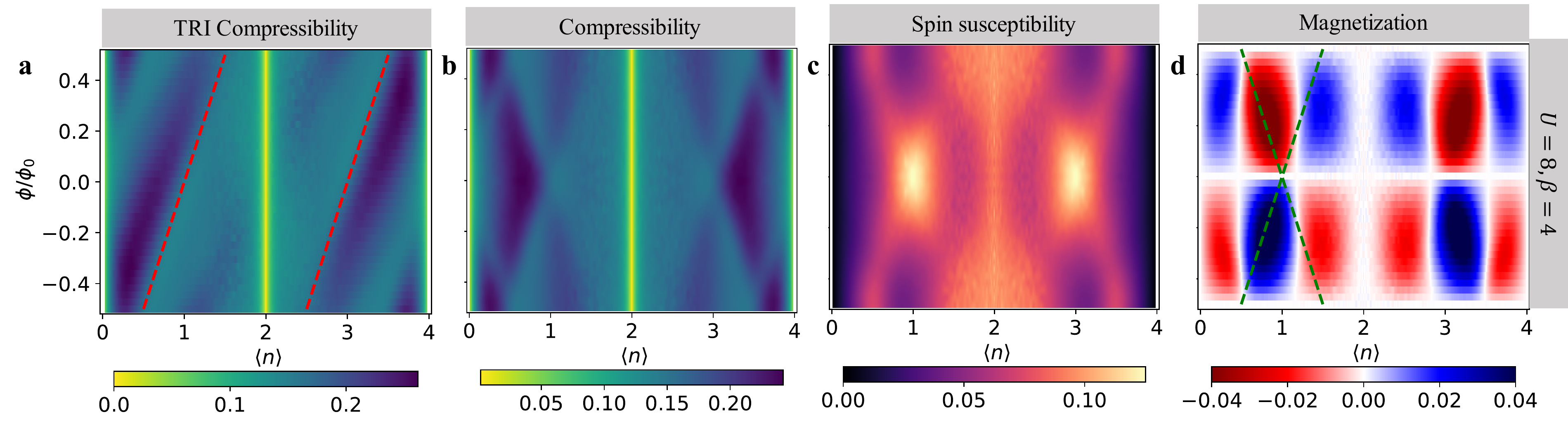}
    \caption{{\bf DQMC simulations of the BHZ-HH model}. DQMC results for the TRI compressibility ({\bf a}), compressibility ({\bf b}), spin susceptibility ({\bf c}) and magnetization ({\bf d}) of the BHZ-HH models at $U/t=8$, $\beta=4/t$. The dashed green lines in panel (d) serves as a guide for the crossing Landau levels signaling the QSH effect.
    }
    \label{fig:BHZH}
\end{figure*}

The simulation results for the BHZ-HH models at $U=8t, \beta=4/t$ are presented in \figdisp{fig:BHZH}. In \figdisp{fig:BHZH}a, two (red) straight lines appear from the zero-field $1/4-$ and $3/4-$filled system, whose inverse slope indicates that the corresponding zero-field $1/4-$ and $3/4-$filled BHZ-HH systems present QSH feature with $C_s=1$ while the $1/2-$filled system becomes a topologically trivial Mott insulator with $C_s=0$. This physics becomes much clearer by studying the standard charge compressibility in \figdisp{fig:BHZH}b which
reveals identical features at $\langle n\rangle=1$ and $\langle n\rangle =3$ of left and right moving LLs indicative of the QSH effect.  Also, the spin susceptibility exhibits a peak  both at $\langle n\rangle=1$ and $\langle n\rangle =3$. The simultaneous appearance of compressibility minima and spin-susceptibility maxima are key features of this Mottness-driven QSH effect, in contrast to its non-interacting counterpart. The magnetization in \figdisp{fig:BHZH}d is also asymmetrical indicating a possible tendency towards ferromagnetism at $\langle n\rangle=1$ and $\langle n\rangle =3$. We return to this in a later section.

 
 To corroborate our findings, we conducted a finite-size analysis (see supplement) and confirm that the same spin Chern number survives in system sizes as large as $N_\text{site}=12\times12$ with insignificant finite-size effects and hence our results are valid in the thermodynamic limit.  We conclude then that the DQMC exhibits the QSH effect at high temperatures at 1/4-filling when $U$ is sufficiently large.   


\subsection*{Exactly solvable model for interacting quantum spin Hall insulators}

The natural question arises: why is $1/4$-filling the new topologically relevant filling and can it be understood in a simple way?  The answer is yes.  For a system with 2 atoms per unit cell, there should be interaction-induced insulating states at any integer filling up to 4 charges in each unit cell. The first such state should be at 1/4-filling.  This physics arises naturally from a momentum-space formulation of the interactions which will result in 4-poles of the Green function, each corresponding to the four insulating states possible.  

We now introduce the Hatsugai-Kohmoto (HK) interaction\cite{HK,HKnp1,HKnp2} into a general QSH Hamiltonian,
\beq
\begin{aligned}
H&=\sum_{{\bf k},\sigma}\big[(\varepsilon_{+,{\bf k},\sigma}-\mu)n_{+,{\bf k},\sigma}+(\varepsilon_{-,{\bf k},\sigma}-\mu)n_{-,{\bf k},\sigma}\big]
\\&+U\sum_{\bf k}(n_{+,{\bf k},\uparrow}n_{+,{\bf k},\downarrow}+n_{-,{\bf k},\uparrow}n_{-,{\bf k},\downarrow}). \label{HHK}
\end{aligned}
\eeq
Without loss of generality, we use the dispersions from the BHZ model (see Methods) setting $M=1$ as an example. This interaction introduces Mottness by tethering double occupancy to k-space rather than the usual real space as in the well-known Hubbard model.  As we will show, this model yields  physics for strong interactions consistent with the Hubbard model.  The reason for this consilience\cite{HKnp2} is that both models break the underlying $\mathbb Z_2$ (distinct from the classification scheme for topological insulators) symmetry of the non-interacting Fermi surface\cite{Anderson}.  As the interaction commutes with the kinetic term, the original non-interacting wave function is untouched and momentum $\bf k$ remains a good quantum number. Therefore, it makes sense to extract the Chern number from an integration over the Brillouin zone.  The interacting Green function can be written down analytically\cite{mfp,HKnp1} as
\beq
\begin{aligned}
G_{\pm,{\bf k},\sigma}(\omega)&=\frac{1-\langle n_{\pm,{\bf k}\bar{\sigma}}\rangle}{\omega+\mu-\varepsilon_{\pm,{\bf k},\sigma}} + \frac{\langle n_{\pm,{\bf k}\bar{\sigma}}\rangle}{\omega+\mu-(\varepsilon_{\pm,{\bf k},\sigma}+U)}.\label{GF}
\end{aligned}
\eeq
The Green function immediately reveals the effect of the correlations. The non-interacting lower and upper bands which were degenerate for spin-up and -down electrons split into singly and doubly occupied sub-bands as a result of Mottness. In the following, we use the abbreviation LSB and LDB for lower singly and doubly occupied sub-bands respectively, and likewise USB and UDB for the upper bands. The energy of the LSB and USB remains at the non-interacting value, while the LDB and UDB move up by a value equal to $U$. For a large enough $U$, the quarter-filled system emerges as an insulator with a filled LSB.  This physics falls out naturally from the HK model because of the 4-pole structure of the Green function.

Since the interaction mixes the spin channels, leading to a huge degeneracy ($d=2^{N_c}$) in the ground state ($N_c$ is the number of unit cells), we need to average over all degenerate ground states\cite{Niu} to rigorously calculate the spin Chern number: $\bar{C}_s=\bar{C}_\uparrow-\bar{C}_\downarrow$. For each spin, the contribution is 
\beq
\bar{C}_{\sigma}= \frac{1}{d}\sum_{\Omega=1}^d \frac{1}{2\pi} \int d^2k f_{xy,\sigma} \langle \Omega|n_{{\bf k},\sigma}|\Omega\rangle, \label{chern}
\eeq
where $f_{xy,\sigma}$ is the normal Berry curvature defined with Bloch wave function \cite{TKNN} because $k$ remains a good quantum number in the HK model, and $(1/2\pi) \int d^2k f_{xy,\sigma}=C_{0\sigma}$. When $U=0$, $\langle \Omega|n_{{\bf k},\sigma}|\Omega\rangle=1$ below the chemical potential. When $U$ is finite, $\langle \Omega|n_{{\bf k},\sigma}|\Omega\rangle$ can be $0$ or $1$. We can conduct the average first for \disp{chern}. When $U$ is large enough to fully separate the singly and doubly occupied bands, $(1/d)\sum_{\Omega=1}^d\langle \Omega|n_{{\bf k},\sigma}|\Omega\rangle=\langle n_{{\bf k},\sigma}\rangle=\langle n_{\sigma}\rangle=1/2$. Then \disp{chern} becomes
\beq
\bar{C}_{\sigma}= \langle n_{\sigma}\rangle \frac{1}{2\pi} \int d^2k f_{xy,\sigma}=\langle n_{\sigma}\rangle C_{0\sigma}=\frac{C_{0\sigma}}{2}. \label{chern2}
\eeq
Thus, the spin Chern number $C_s=C_{0s}/2$ (we will drop the average bar symbol in the following text.). This result demonstrates that each momentum state is equivalently occupied by half spin-up and half spin-down electrons on average. Similarly, the LDB has the same $C_s$, while the USB and UDB have the opposite $C_s$. In short, the strongly correlated quarter-filled system becomes a Mott insulator with a spin Chern number $C_s=C_{0s}/2$ should the interaction exceed the bandwidth. 

\begin{figure}[ht]
    \centering
    \includegraphics[width=0.8\textwidth]{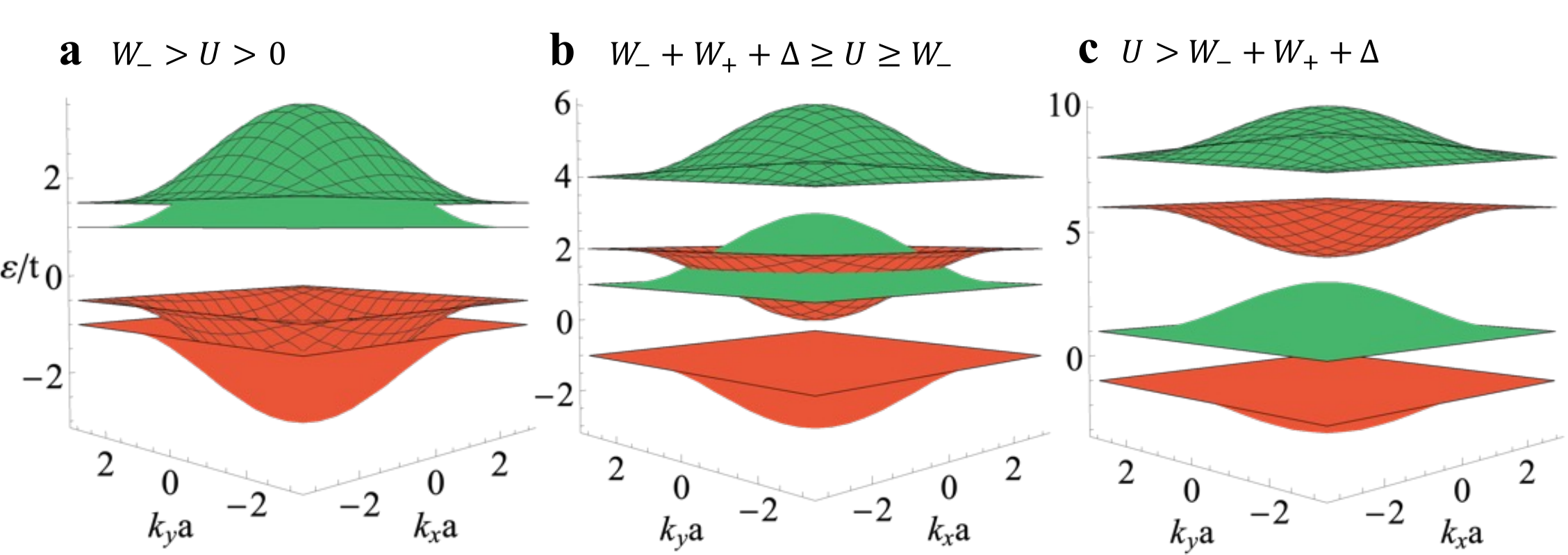}
    \caption{ {\bf Band structure for BHZ-HK model in \disp{HHK} with $M=1$}. Different phases emerge as $U$ increases:  {\bf a} $1/2$-filled QSH insulator for $W_{0-}>U>0$ ($U=0.5$), {\bf b} $1/4$-filled TMI and $1/2$-filled metal for $W_{0-}+W_{0+}+\Delta_0\geq U\geq W_{0-}$ ($U=3$), and {\bf c} $1/4$-filled QSH Mott insulator and $1/2$-filled topologically trivial Mott insulator for $U>W_{0-}+W_{0+}+\Delta_0$ ($U=7$). The red (or green)  color represents $C_s=1$ (or $ -1$). The unmeshed (meshed) band consists of only singly (doubly) occupied states.  It is the splitting of these bands by the interaction that gives rise to the Mott-derived topological physics.
    }
    \label{fig:BHZHK}
\end{figure}

To visualize how this phase emerges,  we plot the band structure in \figdisp{fig:BHZHK} for varying $U$. With $M=1$, the bandwidth for the lower and upper BHZ bands is $W_{0+(-)}=2$ and $\Delta_0=2$ is the topological gap. The non-interacting lower band has $C_{0s}=2$, while the upper band carries the opposite spin Chern number. We separate the non-interacting lower and upper bands into LSB (red-unmeshed), LDB (red-meshed), USB (green-unmeshed) and UDB (green-meshed). As derived above, the red and green sub-bands have the spin Chern number $C_s=1$ and $-1$ respectively. Turning on the interaction causes the doubly occupied sub-bands to increase in energy while the singly occupied sub-bands remain unchanged. For small interactions $W_{0-}> U>0$ (\figdisp{fig:BHZHK}a), the band structure only slightly departs from the non-interacting case. As $U$ increases to $W_{0+}+W_{0-}+\Delta_0 \geq U \geq W_{-}$ (\figdisp{fig:BHZHK}b), the same-color sub-bands fully separate, leading to a gap opening at quarter-filling. Then both the $1/4$- and $3/4$-filled systems become a TMI with a spin Chern number $C_s=1$, while the $1/2$-filled case becomes a conductor. Upon further increasing $U$ to $U>W_{0+}+W_{0-}+\Delta_0$ (\figdisp{fig:BHZHK}c), the $1/2$-filled state becomes a topologically trivial Mott insulator.  All the while, the QSH Mott insulator at $1/4$- and $3/4$-fillings persists with a gap equal to $\Delta$. For a different $M$, the intermediate panel b may change, while panel c is always valid for a large enough $U$. This indicates that generally in the presence of strong interactions, the system becomes a QSH Mott insulator at $1/4$- and $3/4$-filling with spin Chern number $C_s=C_{0s}/2$ and a trivial Mott insulator at $1/2$-filling.   


As we compute in the supplement, the spin susceptibility for the HK model diverges at $T=0$ indicating that the HK model is unstable to ferromagnetic order in this limit. Note that this conclusion applies to a general QSH Hamiltonian (not only to the BHZ model) with HK interactions. This result is consistent with the divergence of the spin-susceptibility of the flat-band KM-Hubbard models. Ultimately this means that the 1/2-filled QSH effect would give rise to a 1/4-filled QAH effect at $T=0$. 

To summarize, this simple model offers a way of understanding why 1/4-filling is special in the Hubbard model.  Note the agreement with the Hubbard simulations is non-trivial because momentum mixing is not present in HK model but is in the Hubbard model.  Hence, the agreement demonstrates that it is the ultimate 4-pole structure of the underlying single-particle Green function that dictates the physics.  As we have shown previously\cite{jpfixed,HKnp2}, the HK model represents a fixed point in which no short-range repulsions are relevant not even Hubbard interactions.  Hence, the HK model is the fixed point for Mott physics.  Possible ferromagnetism at $T=0$ would eventually turn the QSH effect into the QAH effect. Hence, as a result of interactions, the QAH effect appears as the symmetry-broken phase of the QSH effect much the way antiferromagnetism is the low-temperature symmetry-broken phase of a Mott insulator.  Equal drivers of this spontaneous symmetry breaking are consistency with the LSM theorem and the restriction that the Chern number must be an integer. As is evident at 1/4-filling, the QAH always dominates as the symmetry-broken ground state. This is the primary conclusion of this work.

In a previous exact diagonalization study\cite{Neupert3} on a strict flat band model with Hubbard interactions and spin-orbit interaction, it was noticed that ferromagnetism emerged at $1/4$-filling.  This result can be viewed as a special case of HK physics because in the strict flat-band limit of the model studied, any value of $U$ will necessarily prohibit double occupancy thereby producing a gapped state.  The HK result is more general than this result as the gap persists even when the bands disperse.



\subsection*{Gap Opening}
In the previous sections, we have shown that both simulations on the Hubbard model and analytical calculations on the HK model indicate the emergence of non-trivial topology at $1/4$-filling driven by strong correlations. In the non-interacting case, the topology appears with a bulk gap. In the strongly correlated case, however, this is not necessarily true. While a gap opens in the HK model as long as $U$ exceeds the total bandwidth, the precise condition for opening a gap in the Hubbard model is much more subtle because of the dynamical mixing between the bands. In the Hubbard case, the interaction strength needs to exceed a critical value ($U_c^{\text{topo}}\gg W_{0-}$) to induce the topology at $1/4$-filling  and a separate critical value ($U_c^{\text{gap}}$) to open a gap. In general, we find $U_c^{\text{topo}}< U_c^{\text{gap}}$.

\begin{figure}[ht]
    \centering
    \includegraphics[width=0.7\textwidth]{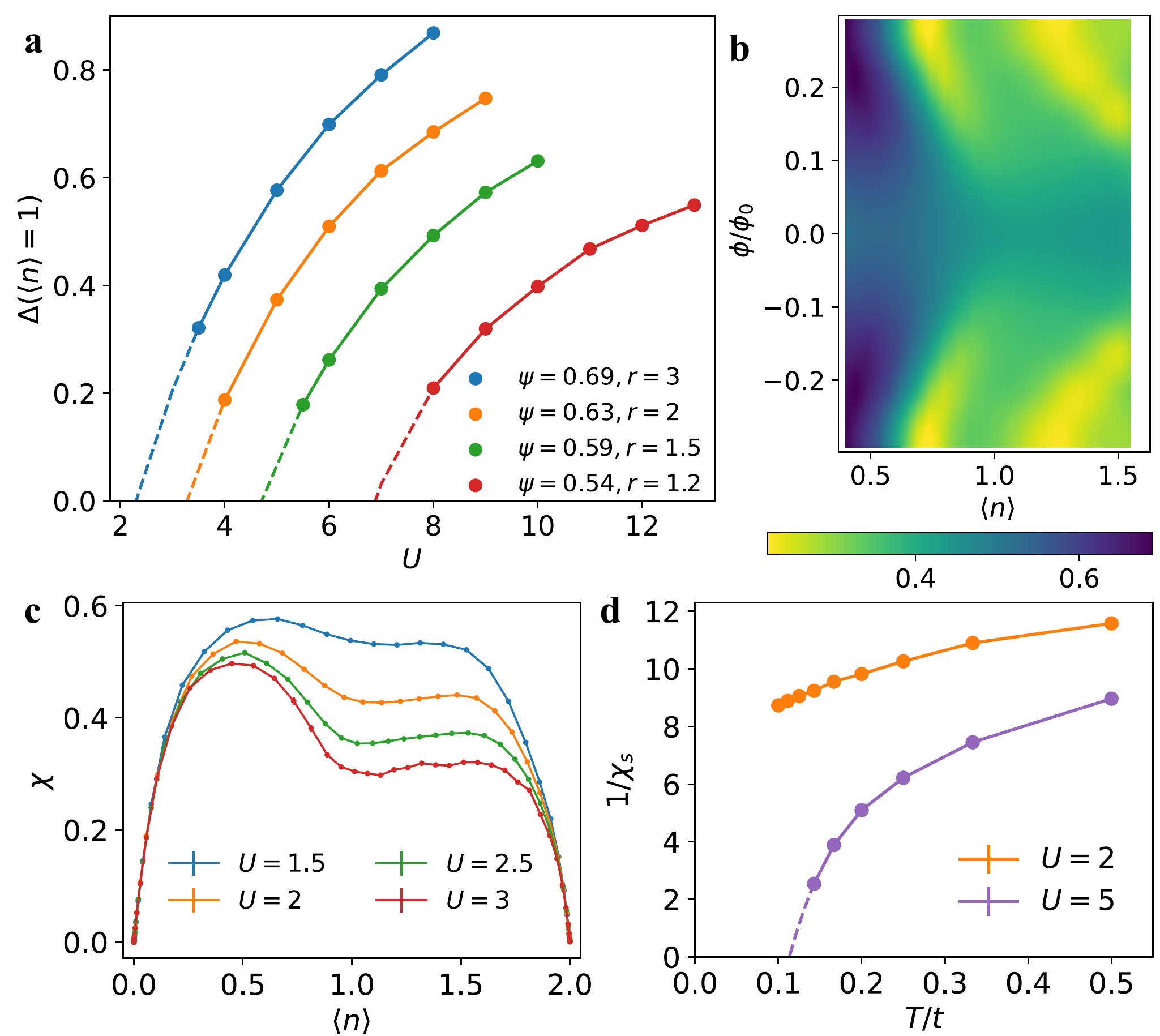}
    \caption{ {\bf Gap opening and non-trivial topology at quarter-filling of the generalized KM-HH model}. {\bf a} Estimated gap of the generalized KM-Hubbard model at quarter-filling as a function of the interaction strength $U$ and the hopping phase $\psi$. These results are obtained from DCA simulations at a temperature of $\beta=20/t$. {\bf b} The DQMC compressibility for the KM-HH model at $U=2t, \beta=8$ around quarter-filling. {\bf c} The zero-field DQMC compressibility for the KM-Hubbard model at various $U$ and $\beta=8$ as a function of the density. {\bf d} The inverse temperature-dependent spin susceptibility from DQMC for KM-Hubbard model at $U=2t$ and $5t$. Panels {\bf b}-{\bf d} fix $\psi=0.63$. All DQMC simulations are done on a $6\times6\times2$ cluster while the DCA simulations are on a $2\times2\times2$ cluster.
    }
    \label{fig:gap}
\end{figure}

From the dip of the high-temperature compressibility computed by DQMC, we can tell roughly when the non-trivial topology appears and hence we are able to extract $U_c^{\text{topo}}$. However, to access the gap information, one has to explore much lower temperatures. This can not be done by DQMC as  we are restricted by the Fermion sign problem and finite-size effects (see supplement for details). To address this problem, we resort to DCA\cite{MaierRMP2005,DCApp,Mai2,Mai3}. We computed the value of the gap defined as
\beq
\Delta(\langle n\rangle=1)=\mu(\langle n\rangle=1.01)-\mu(\langle n\rangle=0.99),
\eeq
in the vicinity of the quarter-filled state in the generalized KM-Hubbard model using DCA  on a $2\times2\times2$ cluster at low temperature $\beta=20/t$. To make contact with previous work on flat-band systems, we define the ratio $r=\Delta_0/W_{0-}$ and study the evolution of the gap as a function of the complex hopping phase, $\psi$ in the generalized KM model (fixing $t^\prime=0.3$). The results are summarized in \figdisp{fig:gap}a. We only plot the data when $\Delta(\langle n\rangle=1)\gtrsim 0.2$ because $\Delta(\langle n\rangle=1)$ by definition remains a small value even when the state is metallic and obtain the $U_c^{\text{gap}}$ by extrapolation to zero gap. In all cases, $\Delta(\langle n\rangle=1)$ is significantly smaller than $\Delta_0$ even when $U>W_{0-}+W_{0+}+\Delta_0$ ($\approx 6$). As the band becomes more dispersive ($\psi$ decreases, or $r$ decreases), $\Delta(\langle n\rangle=1)$ reduces and $U_c^{\text{gap}}$ grows as shown in \figdisp{fig:gap}a ($\Delta_0=2$ for all cases). Now we consider the relation between $U_c^{\text{gap}}$ and $U_c^{\text{topo}}$. Take $\psi=0.63$ as an example ($\Delta_0=2, W_{0-}=1$). Already at $U=2t$, the corresponding KM-HH model shows QAH topology at $\beta=8$ in \figdisp{fig:gap}b, while $U_c^{\text{gap}}\approx3.25$. The DQMC compressibility at zero field is shown in \figdisp{fig:gap}c at various $U<U_c^{\text{gap}}$. It exhibits that a dip at $\langle n \rangle=1$ starts to develop (and therefore the topological magnetic response) at a smaller $U$ before the gap actually opens (see supplement for a benchmark between DCA and DQMC). This supports a topological Mott semimetal (TMSM). In \figdisp{fig:gap}d, we find that for the semi-metallic state, the inverse spin susceptibility $1/\chi_s$ decreases slowly with temperature and is unlikely to reach 0 at finite temperatures, while for the insulating state, the $1/\chi_s$ drops much sharper with temperature so that its extrapolation supports a finite-temperature transition. We conclude that while the Chern numbers remain the same in the TMSM and insulating QAH phases, the TMSM phase lacks a gap and ferromagnetism as well.

In \figdisp{fig:gap}a, at any finite value of $r>1$, $U$ must exceed a critical value $U_c^{\text{gap}}$ for the gap to form. In general, $U_c^{\text{gap}}$ increases as $r$ decreases. When $r=1$ ($\psi=0.5$), we find the gap barely opens for $U<12$ ($\Delta(\langle n \rangle=1, U=12)=0.2$). This situation is directly applicable to the BHZ-Hubbard model because in the topologically relevant region ($-2<M<2$), $r\leq 1$. At $M=1$ (which means $r=1$), $\Delta(\langle n \rangle=1, U=12)=0.22$. However, we already observe the emergence of non-trivial topology at $1/4$- and $3/4$-filling in \figdisp{fig:BHZH} at $U=8$ and even $U=6$ (see supplement). Thus, such phases with non-trivial topology in actuality are semi-metals. Similar to the KM-Hubbard model, the spin susceptibility for these semi-metallic states only shows a soft peak and their temperature evolution doesn't support a finite-temperature transition (see supplement). In such a system, the Hall conductance will have finite temperature corrections \cite{Cooper1} and hence deviate from the value dictated by the Chern number. This observation is in contrast with the recent exact diagonalization study on the BHZ-Hubbard model (with system size up to $3\times4$) which observes gap opening and ferromagnetic order for $U>4$. As we show in the supplement, the finite size effects are sizeable for the cluster size used in this study.  

\subsection*{Experimental Realization}

While the interactions in ultracold atoms in optical traps\cite{TARRUELL2018365} can be adjusted to mimic the physics here, the most obvious synergy is with the moir\'e TMD experiments\cite{WenjinZhao,TingxinLi,Taozui} discussed previously. Our DQMC simulation result in \figdisp{fig:KMHflat} for the flat-band KM-HH model is consistent with this experiment in the existence of QAH and QSH at $1/4$- and $1/2$-filling, respectively.

\begin{figure}[ht]
    \centering
    \includegraphics[width=0.7\textwidth]{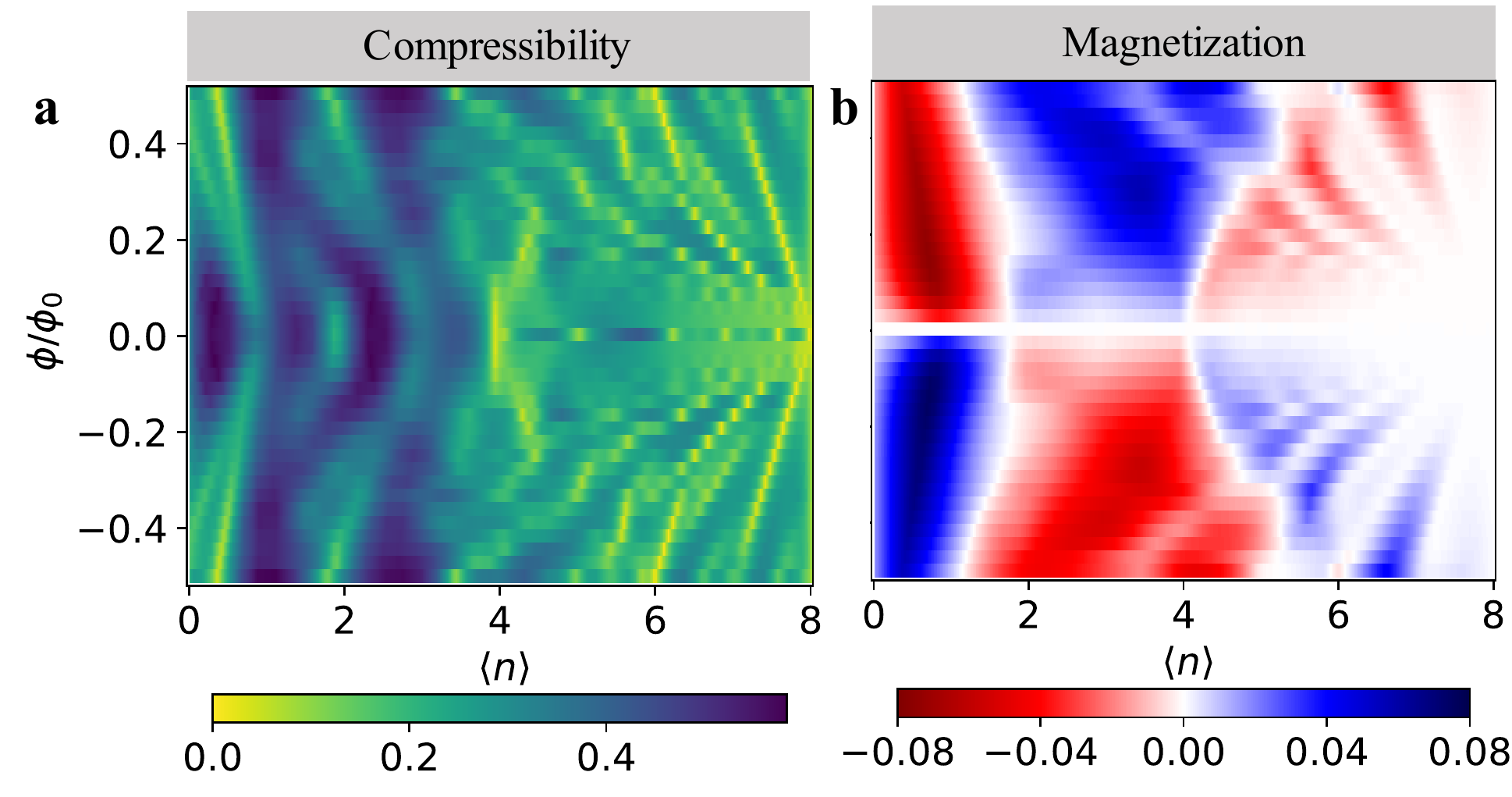}
    \caption{{\bf Compressibility and magnetization of the bilayer KM-HH model}. DQMC results for the bilayer KM-HH model at $U=1.5t, \beta=12/t$ with an interlayer hopping $t_\perp=0.3t$ and voltage difference between the two layers of $V=0.4t$. Panels {\bf a} and {\bf b} show the compressibility and magnetization, respectively, as a function of magnetic flux and electron density.
    }
    \label{fig:KMHbilayer}
\end{figure}

However, we cannot make direct contact with the observation of valley coherence\cite{Taozui} within a single-layer KM model in which spin-valley locking obtains.  Note relaxing the spin-valley locking constraint of the KM model by reversing the spins in one of the bands relative to the other, as indicated in the experiment\cite{Taozui} (see supplement), would lead to a contradiction with a non-zero Chern number per spin in the band insulator limit.  That is, the moir{\'e} band structure of AB stacked MoTe$_2$/WSe$_2$ bilayer can not be captured by a strict four-band model such as the KM model.  The remedy is to construct an eight-band model (details in supplement) consisting of two copies of the KM model, one for each layer with an effective voltage difference between the layers.  For completeness, we recomputed the compressibility for the bilayer flat-band KM-HH model at an intermediate $U=1.5t$.  Clearly shown in \figdisp{fig:KMHbilayer}(a) is the QAH at $\langle n\rangle =1$, the QSH at $\langle n\rangle=2$ and $\langle n\rangle=4$. Besides, there is also an emergent QAH state at $\langle n\rangle=3$. This prediction has been confirmed in a recent experiment on a moir\'e TMD material\cite{foutty2023mapping}.  

The accompanying magnetization in \figdisp{fig:KMHbilayer}(b) is also consistent with these assignments.  Within the eight-band model, spin polarization requires layer coherence because the interaction does not commute with the interlayer hopping and since the same spin is assigned to different valleys in each layer, layer coherence necessarily entails valley coherence. Hence, a simple two-layer extension of our results is sufficient to account for the QAH effect in TMD moir\'e systems.  This reasoning motivates first-principle calculations to determine how the 8-band model should be tailored to apply to specific moir\'e materials.


\section*{Discussion}

Interactions play a non-trivial role in topology in two distinct ways.
First, they lead to a TMSM/TMI with a high-temperature QSH effect characterized by a spin Chern number of $C_s=1$ at quarter filling in the interacting BHZ and KM models. We use the term "Mott" because it is the interactions that lead to a lifting-up of the doubly occupied sector thereby exposing the topology of the 1/4-filled band.  The resultant $C_s=1$ poses a problem as this number must be even for a non-degenerate ground state\cite{barry} with time-reversal symmetry.  The resolution of this dilemma lies in the divergence of the spin susceptibility in the HK model at zero temperature and in the Hubbard model at finite temperature. Both of these indicate a possible spontaneous ferromagnetic phase at $T=0$. The onset of ferromagnetism results with a unit Chern number indicative of the QAH effect and would offer a route around the LSM restriction\cite{lsm2,lsm1} that a unique featureless gapped ground state is impossible  with an odd number of fermions per unit cell. Consequently, our results point to a fundamental reason why the QSH effect at high temperatures must resort to the QAH effect as temperature decreases if a gap opens.  Namely, while at high temperatures, a paramagnetic symmetry-unbroken state obtains, for the ground state to be unique,  the symmetry must be spontaneously broken.  We refer to this onset of the symmetry-broken state as a consequence of topological Mottness\cite{barry2}, in direct analogy with the traditional Mott state which has an antiferromagnetic ground state. 
Therefore, we argue that the $1/4$-filled state is a TMI or TMSM.  In analogy with the traditional Mott insulator with an antiferromagnetic ground state, a TMI exhibits the QSH phase which turns into the symmetry-broken QAH at low temperature as illustrated in Table\ref{table1}. 
 A TMI is qualitatively distinct from the fractional topological insulators\cite{Neupert2,bz2006,Levin,Repellin,Qi2} driven by at least nearest-neighbor interactions. The fractional topological insulator usually consists of two decoupled fractional Chern insulators with opposite spins. However, in the TMI, spin-up and -down electrons are correlated to form the inseparable singly occupied states giving rise to the high-temperature QSH feature and a QAH ground state.  Second, we showed that in the flat-band limit, a high-temperature QAH state exists also at 1/4-filling with an intermediate $U$. In the double-layer extension of this model, the QAH state exhibits valley coherence as is seen experimentally in moir\'e TMD materials\cite{Taozui}.

\section*{Methods} 

\noindent{\bf The Model}. All QSH models are based on Hamiltonians of the form,
\beq
H_{\text{QSH}}=\sum_{\bf k}\Phi^\dagger({\bf k})
\begin{pmatrix}
h_{\text{QAH}}({\bf k}) & 0\\
0 & h_{\text{QAH}}^*(-{\bf k}) \label{QSH1}
\end{pmatrix}\Phi({\bf k}),
\eeq
where $\Phi^\dagger = \{c^\dagger_{O_1,\uparrow} c^\dagger_{O_2,\uparrow},c^\dagger_{O_1,\downarrow} c^\dagger_{O_2,\downarrow} \}$ is a four-component spinor, where $O_{1/2}$ stands for different orbitals or sub-lattices, respectively. \disp{QSH1} means that the spin-up and -down electrons are described by a QAH Hamiltonian $h_{\text{QAH}}({\bf k})=h_a({\bf k})\tau^a$ ($\tau^a$ is the Pauli matrix for orbital/sublattice space) and its TR conjugate counterpart $h_{\text{QAH}}^*(-{\bf k})$ with opposite chirality. As a result, the system is TR invariant and the half-filled case can be a topologically trivial and non-trivial insulator, categorized by a $\mathbb{Z}_2$ invariant or the spin Chern number $C_s$ if $\hat{S}_z$ is conserved.  As a consequence, any ferromagnetism here will be of the Ising type rather than  $U(4)$ as in the BM model\cite{Xie,Bultinck}.  To introduce Hubbard on-site interactions, we need to resort to a real-space representation of the QSH model.  For concreteness, consider the generalized KM model\cite{WenjinZhao} in the honeycomb lattice under an external magnetic field, namely the KM-Hubbard-Hofstadter (KM-HH) model:
\beq
\begin{aligned}
\label{Eq:KMHubbardBfield}
    H&= -\sum_{{\bf i}{\bf j}\sigma} t_{{\bf i},{\bf j}}\exp(i \phi_{{\bf i},{\bf j}} ) c^\dagger_{{\bf i}\sigma}c^{\phantom\dagger}_{{\bf j}\sigma} 
    -\mu\sum_{{\bf i},\sigma} n_{{\bf i}\sigma}
    \\&+ \lambda_\nu(\sum_{{\bf i}\in\text{A},\sigma}n_{{\bf i}\sigma}-\sum_{{\bf i}\in\text{B},\sigma}n_{{\bf i}\sigma})+ U\sum_{{\bf i}}(n_{{\bf i}\uparrow}-\frac{1}{2})(n_{{\bf i}\downarrow}-\frac{1}{2}),
\end{aligned}
\eeq
where $t_{{\bf i},{\bf j}}$ contains the nearest-neighbor hopping $t=1$ (as the energy scale) and next-nearest-neighbor hopping $ t^\prime \text{e}^{\pm i\psi\sigma}$ as the spin-orbit coupling with $\pm i\psi$ following the convention in the Haldane model\cite{Haldane}. If we set $\psi=0.5$ (in the unit of $\pi$), the hopping term reduces to the original KM model\cite{kanemele1,kanemele2}. $\lambda_\nu$ is the sub-lattice potential difference. For simplicity, we fix $\lambda_{\nu}=0$ for this study. Non-trivial topology arises as
long as $t^\prime\neq0$, $\psi\neq0,1$. The phase factor $\exp(i \phi_{{\bf i},{\bf j}})$ which arises from the standard Peierls substitution contains the effect of the external magnetic field, which is introduced to measure the magnetic response of the incompressible states at high temperature to determine the topology. Here $\phi_{{\bf i},{\bf j}}=(2\pi /\Phi_0) \int_{r_{\bf i}}^{r_{\bf j}} {\bf A}\cdot d{\bf l}$, where $\Phi_0=e/h$ is the magnetic flux quantum, the vector potential ${\bf A}=(x\hat{y}-y\hat{x})B/2$ (symmetric gauge), and the integration is along a straight-line path.

The other model we study is the BHZ-Hofstadter-Hubbard (BHZ-HH) model:
\beq
\begin{aligned}
\label{Eq:BHZHubbardBfield}
    H&= t\sum_{{\bf i},\sigma} [ \exp(i \phi_{{\bf i},{\bf i}+\hat{x}} )c^\dagger_{{\bf i},\sigma}\frac{\tau_z-i\sigma \tau_x}{2} c^{\phantom\dagger}_{{\bf i}+\hat{x},\sigma} 
    \\& + \exp(i \phi_{{\bf i},{\bf i}+\hat{y}} )c^\dagger_{{\bf i},\sigma}\frac{\tau_z-i\tau_y}{2} c^{\phantom\dagger}_{{\bf i}+\hat{y},\sigma}
    + \mathrm{h.c.}   ]-\mu\sum_{{\bf i},\sigma} n_{{\bf i},\sigma} 
    \\&+M\sum_{{\bf i},\sigma}c^\dagger_{{\bf i},\sigma}\tau_z c_{{\bf i},\sigma} 
    + U\sum_{{\bf i}\alpha}(n_{{\bf i}\alpha\uparrow}-\frac{1}{2})(n_{{\bf i}\alpha\downarrow}-\frac{1}{2}),
\end{aligned}
\eeq
where $t=1$ (energy scale), $\tau_a$ is the Pauli matrix in the orbital basis and $\alpha$ is the orbital index. Non-trivial topology arises as long as $|M|<2$.  

At zero field, a general QSH Hamiltonian can be diagonalized into 
\beq
H_{\text{QSH}}=\sum_{{\bf k},\sigma}\big[(\varepsilon_{+,{\bf k},\sigma}-\mu)n_{+,{\bf k},\sigma}+(\varepsilon_{-,{\bf k},\sigma}-\mu)n_{-,{\bf k},\sigma}\big],\label{QSH2}
\eeq
where $\mu$ is the chemical potential and
\beq
\varepsilon_{\pm,{\bf k},\sigma}=h_{0,\sigma}({\bf k})\pm\sqrt{h_{x,\sigma}^2({\bf k})+h_{y,\sigma}^2({\bf k})+h_{z,\sigma}^2({\bf k})}
\eeq
represents the upper ($+$) and lower ($-$) bands for each spin. In the BHZ\cite{bhz} model,
\beq
\begin{aligned}
&h_{0,\sigma}({\bf k})=0,~~h_{x,\sigma}({\bf k})=\sigma t \sin(k_x),
\\&h_{y,\sigma}({\bf k})=t\sin(k_y),~~h_{z,\sigma}({\bf k})=M+t\cos(k_x)+t\cos(k_y),\label{BHZ}
\end{aligned}
\eeq
The spin-up and -down electrons have the same dispersion but different wave functions with opposite chirality. For $2>M>0$ (or $-2<M<0$), the half-filled system is a QSH insulator\cite{bhz} with $C_{0s}= 2$ (or $-2$) related to the spin Hall conductance\cite{bhz,Qi1}. 

\noindent{\bf Numerical simulations}. 
We use the DQMC method\cite{Blankenbecler,Hirsch,White} to simulate the KM-HH and BHZ-HH models on an $N_{\text{site}}=6\times6\times2$ cluster (two sublattices or orbitals per unit cell) with modified periodic boundary conditions\cite{assaad}. A single-valued wave function requires the flux quantization condition $\Phi/\Phi_0=n_f/N_c$ (with $n_f$ an integer). We also use DCA to calculate the charge gap at low temperatures on a $N_{\text{site}}=2\times2\times2$ cluster of the KM-MM model. The DCA represents the infinite lattice in the thermodynamic limit by a finite cluster embedded in a self-consistent dynamical mean field. It has a much milder finite-size effect and Fermion sign problem. The technical details of these two methods are provided in the supplement.

\section*{Data Availability} 
The DQMC and DCA data generated in this study have been deposited in the Zenodo under the accession code \url{https://doi.org/10.5281/zenodo.8275156}.
\\

\section*{Code Availability} The DQMC code used for this project can be obtained at \url{https://doi.org/10.5281/zenodo.8275145}. The DCA code for this study can be obtained at \url{https://doi.org/10.5281/zenodo.8275154}. 
\\

\section*{Acknowledgements} We thank Taylor L. Hughes, Edwin W. Huang, Kin Fai Mak and Kam Tuen Law, Cristian Batista, Thomas Maier, Charlie Kane and Barry Bradlyn for useful discussions. We also thank P. Armitage for help with the pithy title.  This work was supported by the Center for Quantum Sensing and Quantum Materials, a DOE Energy Frontier Research Center, grant DE-SC0021238 (P.M. B.E.F., and P.W.P.). PWP also acknowledges NSF DMR-2111379 for partial funding of the HK work which led to these results. The DQMC calculation of this work used the Advanced Cyberinfrastructure Coordination Ecosystem: Services \& Support (ACCESS) Expanse supercomputer through the research allocation TG-PHY220042, which is supported by National Science Foundation grant number ACI-1548562\cite{xsede}.
\\

\section*{Author Contributions} 
P.M. performed the DQMC and DCA calculations on the Hubbard model, analyzed the data, and carried out the analytic calculations on the HK model; J.Z. provided the method to calculate the spin Chern number in HK model and analyzed the data; B.E.F. provided information on experimental realizations; P.W.P. supervised the project; P.M. and P.W.P. wrote the paper with input from all authors. 
\\

\section*{Competing interests}
The authors declare no competing interests. 
\\

 \begin{table}[ht]
\begin{center}
\begin{tabular}{ | p{2cm} | p{1.5cm}| p{1.5cm} | p{1.5cm}| p{1cm} | p{1cm}| } 
  \hline
 \multicolumn{6}{| c |}{\bf Classification Scheme} \\
  \hline
   &  & TMI & TI & MI & BI  \\
   \hline
     Filling &  & $1/4$ & $1/2$ & $1/2$ & $1$ \\ 
  \hline
  Topology & $T>T_\text{th}$ & QSH & QSH  & trivial & trivial \\ 
     &  &  ($C_s=1$) &  ($C_s=2$) &  &  \\ 
  \cline{2-6}
   & $T=0$ & QAH  & QSH & trivial & trivial \\ 
      &  &  ($C=\pm1$) &  ($C_s=2$) &  &  \\ 
  \hline
  Magnetism & $T>T_\text{th}$ & PM & PM & PM & PM \\ 
  \cline{2-6}
   & $T=0$ & FM & PM & AFM & PM \\
  \hline
\end{tabular}  
\end{center}
  \caption{ {\bf The comparison between four different insulators}. TMI: topological Mott insulator; TI: topological insulator; MI: Mott insulator; BI: band insulator. PM: paramagnetism; FM: ferromagnetism; AFM: antiferromagnetism. $T_\text{th}$ is a threshold temperature above which the symmetry is maintained. Here we assume two atoms per unit cell for all cases, and $U$ is sufficiently large for Mott physics to dominate in a TMI and MI. } \label{table1}
\end{table}

\end{document}


\title{Supplementary information}
\author{ Peizhi Mai$^{1}$, Jinchao Zhao$^{1}$, Benjamin E. Feldman$^{2,3,4}$
 and Philip W. Phillips$^{1,\dagger}$ }

\affiliation{$^1$Department of Physics and Institute of Condensed Matter Theory, University of Illinois at Urbana-Champaign, Urbana, IL 61801, USA}
\affiliation{$^2$Geballe Laboratory of Advanced Materials, Stanford, CA 94305, USA}
\affiliation{$^3$Department of Physics, Stanford University, Stanford, CA 94305, USA}
\affiliation{$^4$Stanford Institute for Materials and Energy Sciences, SLAC National Accelerator Laboratory, Menlo Park, CA 94025, USA}

\date{July 2022}


\maketitle
\section{Overview}
In this supplement, we provide further information to support our conclusion for the interaction-driven topological Mott semimetal (TMSM) and insulator (TMI) with quantum anomalous Hall (QAH) effect in a general strongly correlated quantum spin Hall (QSH) system. In the following, we first include the details of determinantal quantum Monte Carlo (DQMC) and dynamical cluster approximation (DCA) simulations. Then we have three sections of supplemental results for the Kane-Mele-Hofstadter-Hubbard (KM-HH) and Bernevig-Hughes-Zhang-Hofstadter-Hubbard (BHZ-HH) from DQMC and DCA simulations as well as Hatsugai-Kohmoto (HK) model from analytical calculations. At the end, we also have a section to discuss how the bilayer flat-band KM-HH model explains and predicts the experiments.

\section{Details of DQMC simulations}
We use the DQMC code in \href{https://github.com/edwnh/dqmc}{https://github.com/edwnh/dqmc}. The Hubbard-Stratonovich transformation is not SU(2) symmetric. We discretize the imaginary time $\beta$ into $L$ slides with $\Delta\tau=0.1$. We conduct 5000 warmup sweeps and 200000 measurement sweeps (10 measurements per sweep) at each Markov chain. We scan the chemical potential from $-10$ to $10$ (about 160 values) to obtain the density dependence of physical quantities. Depending on the sign problem in \figdisp{fig:sup_KMsign} for example (little change under finite magnetic field), we use different numbers (from $2$ to $40$) of Markov chains to bring down the error bar. 

\section{Details of DCA simulations}
We use the DCA++ package\cite{DCApp} in \href{https://github.com/CompFUSE/DCA}{https://github.com/CompFUSE/DCA}. We use the continuous-time, auxiliary-field quantum Monte-Carlo algorithm\cite{GullEPL2008, gull_submatrix_2011} as the cluster solver for DCA simulations. The expansion order of the CT-AUX QMC algorithm is typically $50$-$200$. Increasing this number can alleviate the sign problem to some extent but that also means doing more measurements effectively. Depending on the sign problem for a given parameter set, we measure $(1\sim4) \times 10^7$ samples for the correlation functions. Six to eight iterations of the DCA loops are typically needed to obtain good convergence for the DCA self-energy and chemical potential for the measurement of the gap.

\clearpage

\section{Supplemental DCA and DQMC results for the KM-HH model}

\subsection{Sign problem of DQMC simulations for the KM-HH model}
\begin{figure}[ht]
    \centering
    \includegraphics[width=0.5\textwidth]{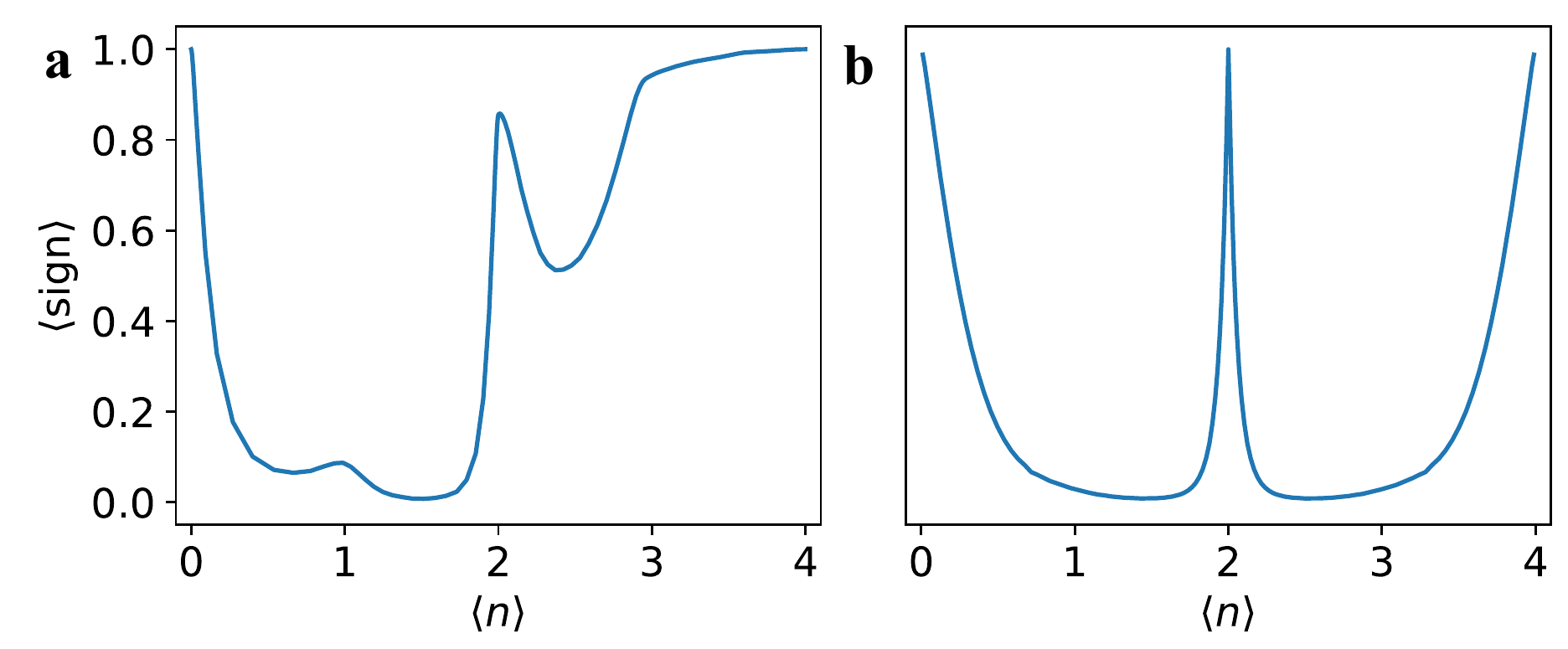}
    \caption{The average sign for the KM-HH model with zero external magnetic flux for {\bf a} $t'=0.3, \psi=0.81, U=3, \beta=7$ and {\bf b} $t'=0.3,\psi=0.5, U=12,\beta=3$.}
    \label{fig:sup_KMsign}
\end{figure}

The DQMC simulation for KM-HH model suffers from severe sign problems in certain doping densities. For this reason, our computation is limited to high temperatures. The average signs for the flat-band case $\psi=0.81$ (in the unit of $\pi$) at $U=3, \beta=7$ and the dispersive case $\psi=0.5$ at $U=12, \beta=3$ as a function of density at the zero field are shown in \figdisp{fig:sup_KMsign}. Note that in the flat-band case (\figdisp{fig:sup_KMsign}a), half-filling is not sign-problem-free due to the lack of particle-hole symmetry. This symmetry is retained in the original KM case (\figdisp{fig:sup_KMsign}b) and thus the average sign is 1 at $\langle n\rangle=2$.

\subsection{Non-interacting results for the KM-HH model at lower temperatures}
\begin{figure}[ht]
    \centering
    \includegraphics[width=0.55\textwidth]{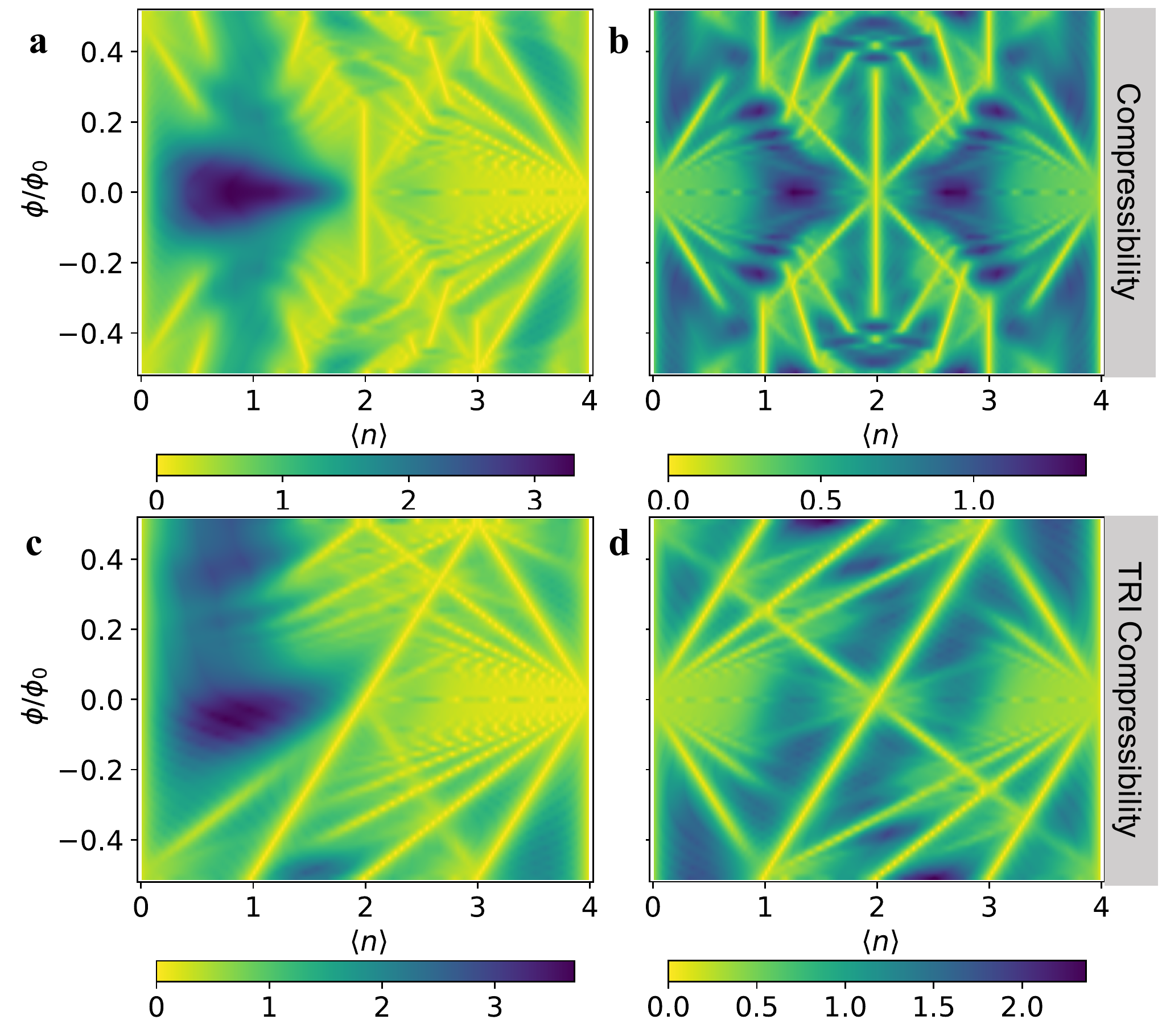}
    \caption{The compressibility (first row) and TRI compressibility (second row) for the non-interacting KM-HH model for ({\bf a},{\bf c}) $t'=0.3, \psi=0.81$ and ({\bf b},{\bf d}) $t'=0.3,\psi=0.5$. The inverse temperature for both cases is $\beta=20$.}
    \label{fig:sup_KMnonintlowT}
\end{figure}

Here we show the non-interacting compressibility and TRI compressibility for the KM-HH model at lower temperature $\beta=20$ in \figdisp{fig:sup_KMnonintlowT} for the flat-band case (a) and the original dispersive case (b). In either case, the integer quantum Hall effects become sharper at this lower temperature. For the TRI compressibility, the inverse slope of the leading middle line crossing $\langle n \rangle=2$ gives the spin Chern number $C_s=2$.
\clearpage

\subsection{The crossover from metal to QAH/QSH features at high temperatures}
\begin{figure}[ht]
    \centering
    \includegraphics[width=0.8\textwidth]{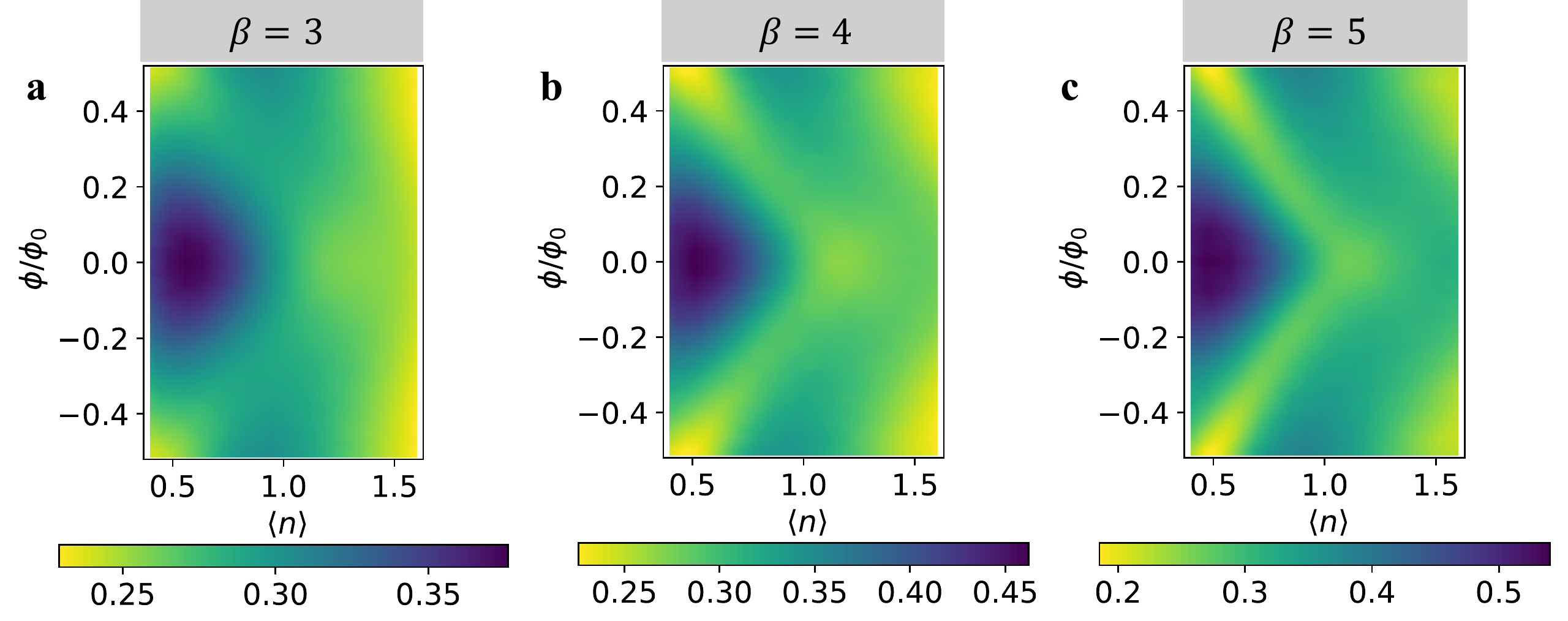}
    \caption{The compressibility for the generalized flat-band KM-HH model ($t^\prime=0.3,\psi=0.81$) as a function of magnetic flux and density with fixed for different temperatures {\bf a} $\beta=3$, {\bf b} $\beta=4$, {\bf c} $\beta=5$.}
    \label{fig:highTcross}
\end{figure}

If the temperature is too high, we would not observe the non-trivial topology emerging at quarter-filling, no matter how large $U$ is. In the Fig.~1 of the main text, at $U=12$, we need to reach $\beta=3$ to discern the dip in the compressibility and hence the QSH feature. If for a smaller $U$, we can expect the onset temperature for such a crossover (from a featureless metal to a high-temperature QAH/QSH effect) to decrease. For example, \figdisp{fig:highTcross} shows that $\beta=4$ is needed to observe the high-temperature QAH behavior at $U=3$.

\subsection{The crossover from QAH to QSH features at high temperatures}
\begin{figure}[ht]
    \centering
    \includegraphics[width=0.8\textwidth]{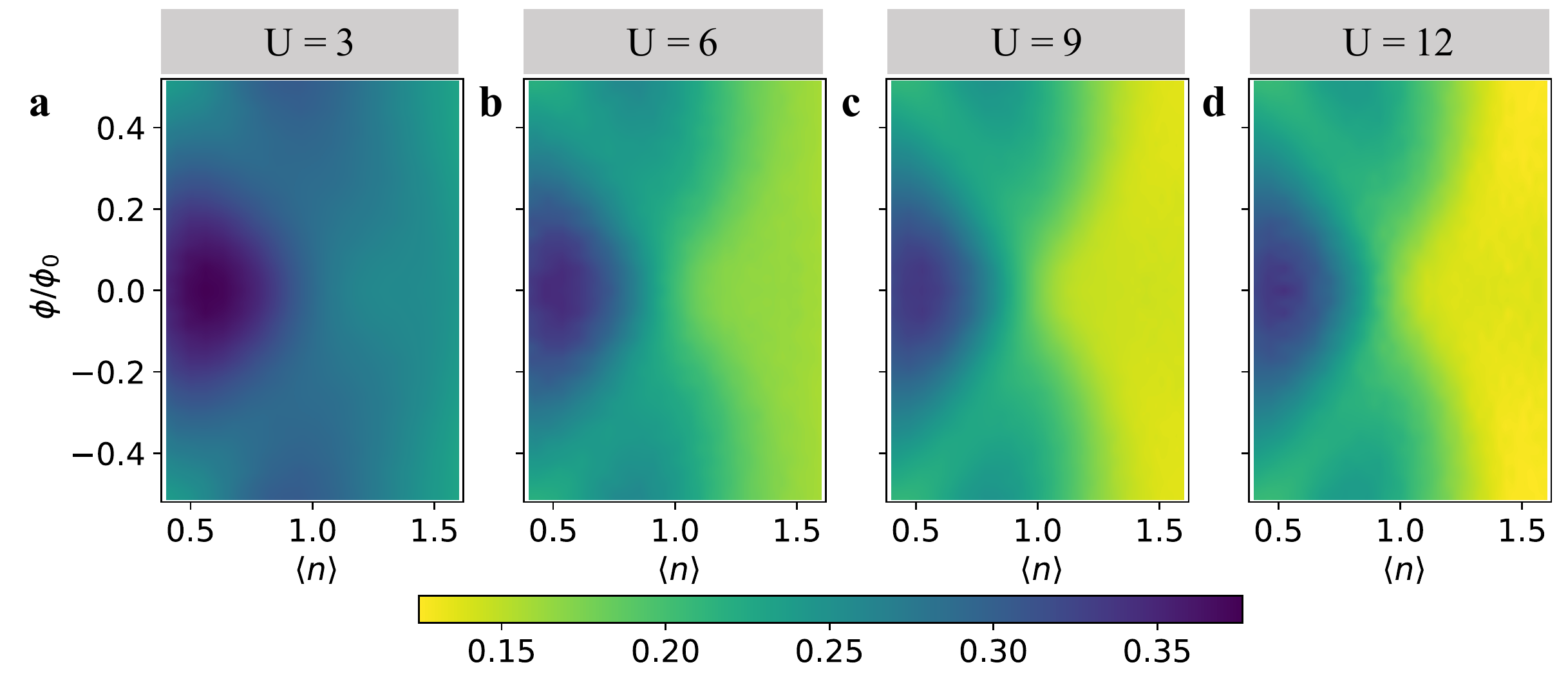}
    \caption{The compressibility for the generalized flat-band KM-HH model ($t^\prime=0.3,\psi=0.81$) as a function of magnetic flux and density at {\bf a} $U=3$, {\bf b} $U=6$, {\bf c} $U=9$ and {\bf d} $U=12$. The inverse temperature for all cases is $\beta=3$.}
    \label{fig:crossover}
\end{figure}

In Fig.~1 of the main text, we jump directly from the QAH effect at $U=3t$ (second row) to the QSH effect at $U=12t$ (third row) and also change the temperature. Here we fill in the gap to see how this happens by gradually increasing $U$ while keeping $\beta=3/t$ in \figdisp{fig:crossover}. As $U$ increases, the right Landau levels start to appear while the left Landau levels become more prominent, suggesting a crossover instead of a transition.

\clearpage

\subsection{The comparison between DCA and DQMC simulations}
\begin{figure}[ht]
    \centering
    \includegraphics[width=0.7\textwidth]{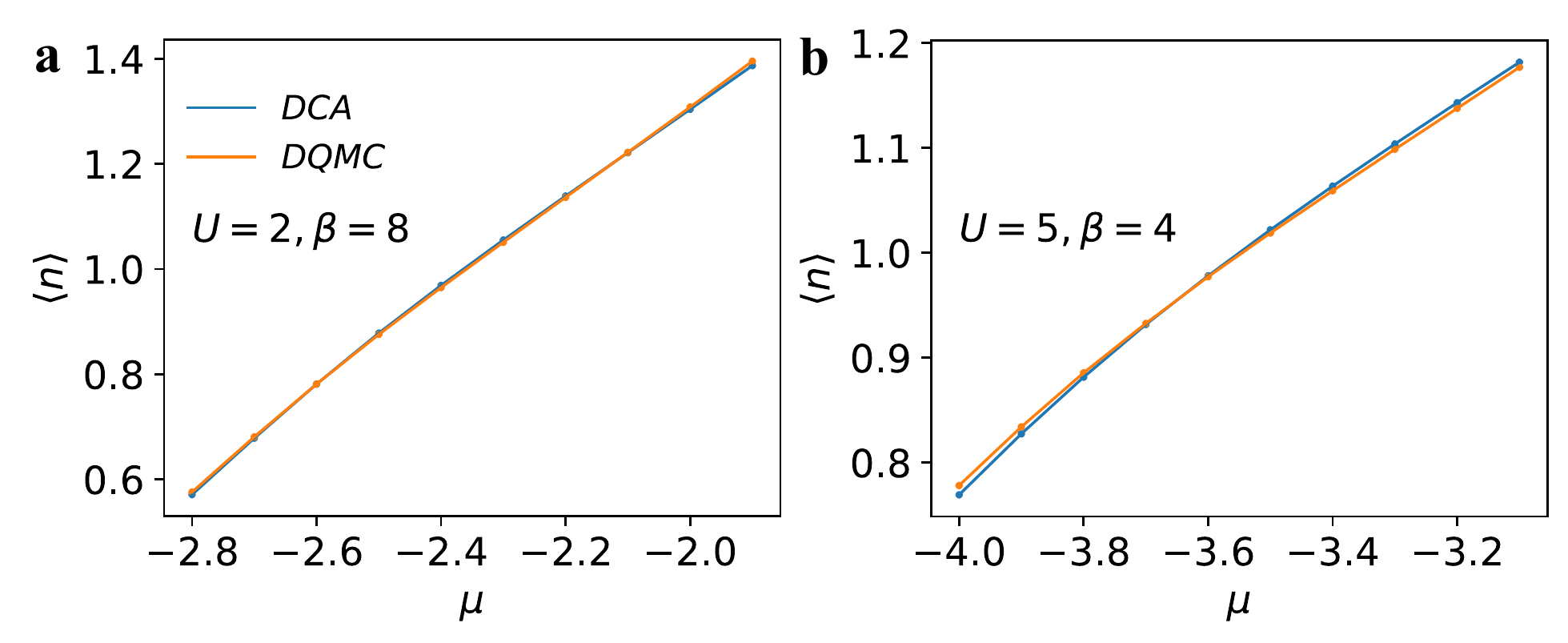}
    \caption{The comparison on $\langle n\rangle$ versus $\mu$ relation between DQMC on a $6\times6\times2$ cluster and DCA on a $2\times2\times2$ cluster at for the generalized KM-Hubbard model ($t^\prime=0.3,\psi=0.63$) at {\bf a} $U=2, \beta=8$ and {\bf b} $U=5, \beta=4$.}
    \label{fig:benchmark}
\end{figure}

In \figdisp{fig:benchmark}, we show a good benchmark between DCA and DQMC on the $\langle n\rangle$ versus $\mu$ relation for the KM-Hubbard model ($t^\prime=0.3,\psi=0.63$) at $U=2,\beta=8$ and $U=5,\beta=4$, supporting Fig.~5 in the main text.
\clearpage

\subsection{DQMC results for the original (dispersive) KM-HH model}
\begin{figure}[ht!]
    \centering
    \includegraphics[width=\textwidth]{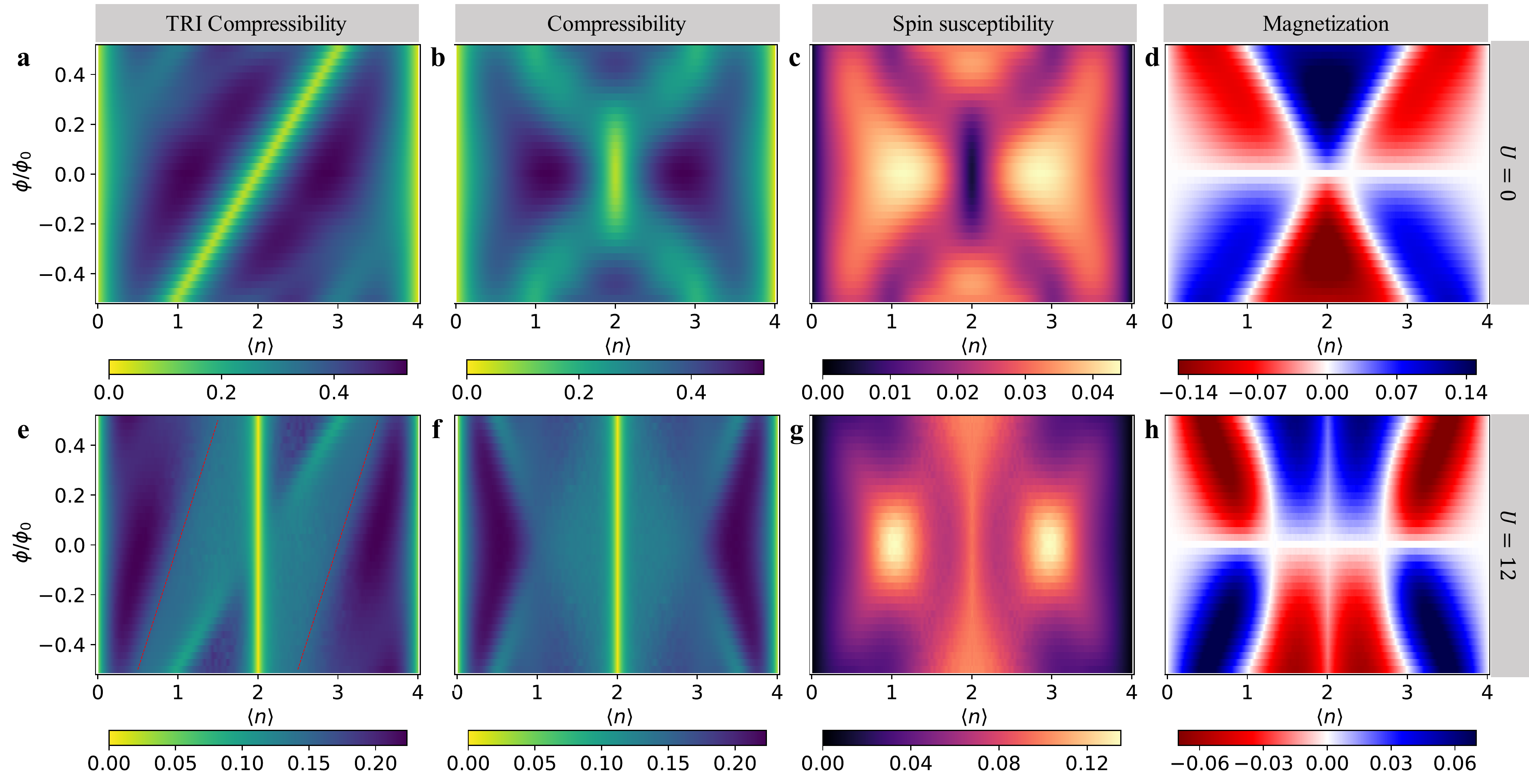}
    \caption{DQMC results for the KM-HH-TRI and KM-HH models at $U=0$ (first row) and $U/t=12$ (second row). The first column shows the compressibility $\chi$ as a function of magnetic flux and electron density for the KM-HH-TRI model. The second to fourth columns show $\chi$, spin susceptibility $\chi_s$ and magnetization $\langle m_z\rangle$ respectively, for the KM-HH model. The temperature is $\beta=3/t$. 
    }
    \label{fig:sup_KMH}
\end{figure}

Here we show the DQMC simulation results in \figdisp{fig:sup_KMH} for the original (dispersive) KM-HH model at $t^\prime=0.3,\psi=0.5$. Comparing panels (a) and (e), we find that the strong correlation induces non-trivial topology at zero-field for $1/4-$ and $3/4-$filling with a "TRI" Chern number $C^{\text{TRI}}=1$, which is akin to that obtained for the spinful Haldane-Hofstader-Hubbard model\cite{mfp}. This means QSH effects emerge at the zero-field $1/4-$ and $3/4-$filled KM-HH systems with a spin Chern number $C_s=1$. The compressibility and spin susceptibility for the KM-HH model in \figdisp{fig:sup_KMH}(f,g) are essentially the same as those for the BHZ-HH model in Fig.~3(b,c) of the main text. In the presence of strong electron-electron interaction, the compressibility displays a pair of zero-mode LLs at quarter-filling where the spin susceptibility has ridges. The relatively noticeable differences between the KM-HH and BHZ-HH models lie in the magnetization at the band edge as is evident from panels \figdisp{fig:sup_KMH}h relative to Fig.~3d in the main text.  Note the sign of the magnetization is not particularly important as a discrepancy already arises with the non-interacting case and hence likely arises from the difference in the lattices. In short, we observe the same QSH effects driven by Mottness in the quarter-filled KM-HH model with large enough Hubbard interaction and adds to the ubiquity of this correlation-driven effect. Unlike the flat-band case, here $W_{-0}=\Delta=2$. There is no such intermediate region of $U$ where a QAH effect emerges at high temperatures as in Fig.~1(d-f). As $U$ becomes sufficiently large, the QSH effect appears.

\clearpage

\section{Supplemental DQMC and DCA results for the BHZ-HH model}

\subsection{Sign problem of DQMC and DCA simulations for the BHZ-HH model}
\begin{figure}[ht!]
    \centering
    \includegraphics[width=0.97\textwidth]{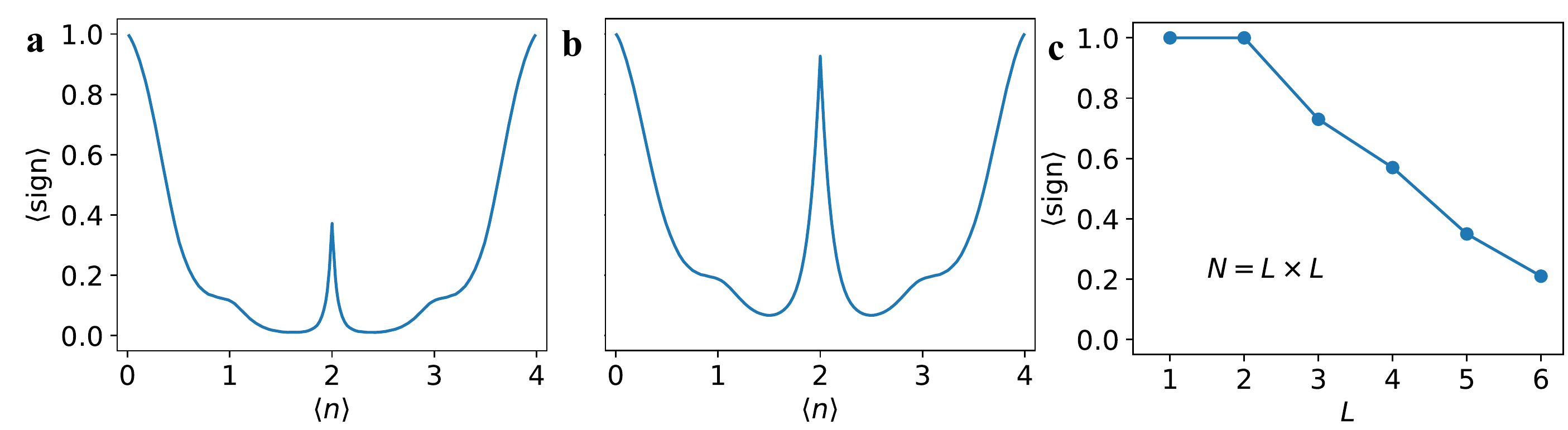}
    \caption{Average sign of DQMC simulations as a function of density for the BHZ-HH model ($M=1$) at zero field and {\bf a} $U=8$,$\beta=4$; {\bf b} $U=12$,$\beta=3$. The DQMC simulation is conducted on a $N=6\times6$ cluster. Panel {\bf c} shows the average sign of DCA simulations as a function of cluster size at half-filling and $U=8,\beta=5$.
    }
    \label{fig:sup_BHZsign}
\end{figure}

Here we calculate the average sign from DQMC simulations for the BHZ-HH model at zero field in \figdisp{fig:sup_BHZsign}(a,b). An interesting observation is that half-filling is not sign-problem-free, unlike the KM-Hubbard case in \figdisp{fig:sup_KMsign}b. Previous studies\cite{Budich,Yoshida2013,Yoshida2012} on the BHZ-Hubbard model at half-filling used dynamical mean field theory (DMFT) and found no sign problem when using quantum Monte Carlo as the impurity solver. To make a connection with these results, we further check the average sign in DCA simulations at different cluster sizes shown in \figdisp{fig:sup_BHZsign}c. DCA is a cluster version of DMFT and DCA reduces to DMFT when the cluster size $N=1$. In \figdisp{fig:sup_BHZsign}c, we find that in the DMFT limit ($L=1$), half-filling is exactly sign-problem-free, agreeing with the previous study. However, its average sign decreases dramatically as $L$ increases (even at $L=2$ the average sign  deviates slightly from 1). Another observation is that if we turn on a staggered sublattice potential $C_\nu$ in the KM-Hubbard model, there would be a sign problem at half-filling despite time-reversal symmetry. Note that all previous quantum Monte-Carlo studies\cite{Hohenadler2012, Hohenadler2011} focused on the half-filled KM-Hubbard model with $C_\nu=0$ and found it to be sign-problem-free.

\begin{figure}[ht!]
    \centering
    \includegraphics[width=0.4\textwidth]{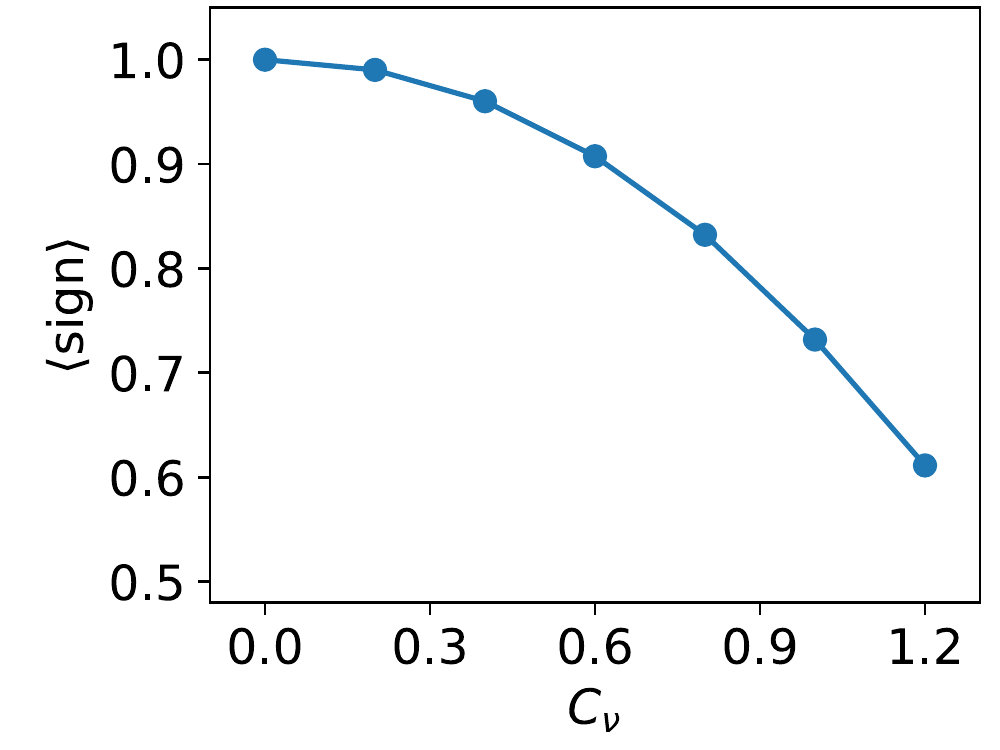}
    \caption{Average sign of DQMC simulations as a function of a staggered sublattice potential for the KM-Hubbard model ($t^\prime=0.3, \psi=0.5$) at $U=8$ and $\beta=4$. 
    }
    \label{fig:sup_KMsign2}
\end{figure}

The explanation for this is as follows. Recall that a famous sign-problem-free example is the half-filled single-band Hubbard model on a square lattice with only nearest-neighbor hopping. This arises because the Hamiltonian is unchanged under a particle-hole transformation $c_{j\sigma}\rightarrow d_{j\sigma}^\dagger(-1)^j$ with $j=\pm1$ for sub-lattice A and B. This particle-hole symmetry is broken if we introduce the next-nearest-neighbor hopping $t'$ which leads to a minus sign under this transformation. In the KM-Hubbard model with $C_v=0$, the Hamiltonian under this transformation is equivalent to flipping a spin ($\uparrow\rightarrow\downarrow, \downarrow\rightarrow\uparrow$). The Hamiltonian is effectively unchanged regarding the sign problem. If $C_v$ is finite, then it acquires a minus sign under this transformation and thus the Hamiltonian becomes different. Therefore, a sign problem is present in \figdisp{fig:sup_KMsign2} for the half-filling KM-Hubbard model when $C_v$ is finite, despite the fact that time-reversal symmetry is still maintained. In the BHZ-Hubbard model, the transformation involving $j=\pm1$ involves different orbitals rather than sub-lattices because each unit cell contains two orbitals. Then under this particle-hole transformation, the diagonal term $M+t\cos(k_x)+t\cos(k_y)$  changes a sign, leading to a change in the Hamiltonian. Due to the lack of this symmetry, there is a sign problem at half-filling as shown in \figdisp{fig:sup_BHZsign} for DQMC simulations. That being said, this problem may be studied using the sign-problem-free quantum Monte Carlo method in Majorana representation\cite{LiZiXiang}. 

\clearpage

\subsection{The analysis of finite size effect and gap opening for the BHZ-HH model with comparison to exact diagonalization}

\begin{figure}[ht!]
    \centering
    \includegraphics[width=0.7\textwidth]{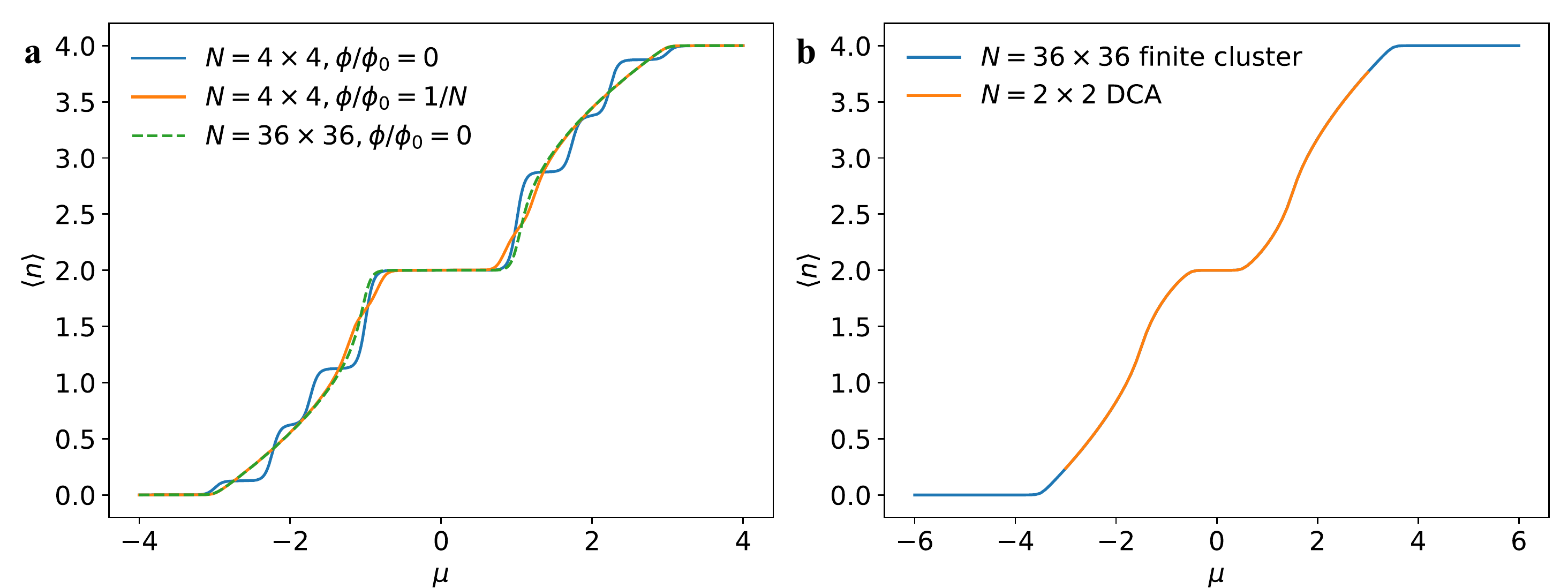}
    \caption{$\langle n\rangle$ as a function of $\mu$ for the non-interacting BHZ-HH models at $\beta=20$ from the calculations on different cluster sizes. {\bf a} The comparison between $N=4\times4$ (under zero and minimal finite field) and $N=36\times36$ under zero field. {\bf b} The comparison between $N=36\times36$ finite cluster calculation and $N=2\times2$ DCA simulations both under zero field. The parameter $M=1$ for panel {\bf a} and $M=1.5$ for panel {\bf b}.
    }
    \label{fig:BHZnonintFSE}
\end{figure}

We conduct DQMC on a $6\times6$ cluster. This small cluster size would suffer finite-size effect at low temperatures. An effective way to overcome it is through a minimal finite magnetic flux. An example of a finite cluster calculation for the non-interacting BHZ-HH model at $\beta=20$ is given in \figdisp{fig:BHZnonintFSE}a. At such a low temperature, the system size has to be as large as $N=36\times36$ to give an accurate finite-cluster description of the single-particle properties. A small  $N=4\times4$ cluster in \figdisp{fig:BHZnonintFSE} would induce unphysical gaps in the $\langle n\rangle$ versus $\mu$ relation. Turning on a minimal magnetic field can provide a much better approximation despite the small effect from the magnetic field, as shown in \figdisp{fig:BHZnonintFSE}a. An alternative way is to conduct a DCA simulation. Since the dynamical mean field approximates the degree of freedom in the infinite bulk lattice outside the cluster, even an $N=2\times2$ DCA cluster can represent the thermodynamic limit in describing the $\langle n\rangle$ versus $\mu$ relation, as shown in \figdisp{fig:BHZnonintFSE}b.

\begin{figure}[ht!]
    \centering
    \includegraphics[width=0.9\textwidth]{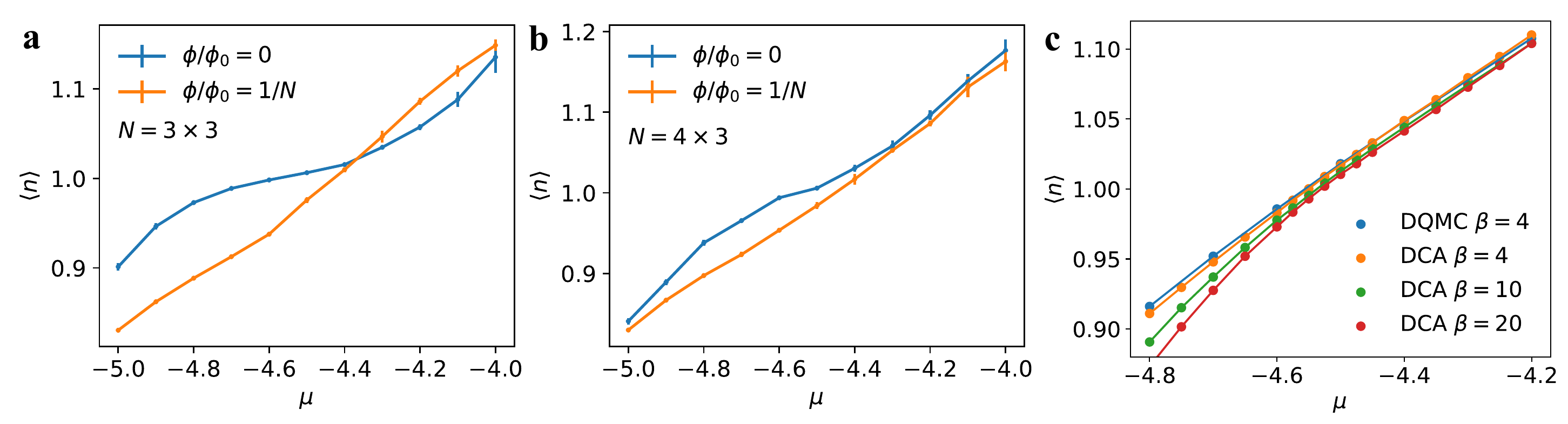}
    \caption{$\langle n\rangle$ versus $\mu$ for the BHZ-HH models ($M=1$) at $U=8$ from DQMC and DCA simulations. Panels {\bf a} and {\bf b} are the DQMC results at minimal magnetic fluxes with $\beta=8$ on a $N=3\times3$ and $N=4\times3$ cluster, respectively. Panel {\bf c} includes DQMC results on an $N=6\times6$ cluster with $\beta=4$ and DCA results on an $N=2\times2$ cluster with $\beta=4,10,20$.
    }
    \label{fig:BHZHFSE}
\end{figure}

Now we turn on the interactions to $U=8$. The recent exact diagonalization study\cite{Yang} is a finite cluster calculation on a system size up to $N=4\times3$ at zero field and zero temperature. DQMC simulations at a relatively high temperature $\beta=8$ for cluster size $N=3\times3$ and $N=4\times3$ already shows considerable finite-size effects as shown in \figdisp{fig:BHZHFSE}(a) and (b) respectively, namely inducing a non-existing gap at $\langle n\rangle=1$. The situation worsens at lower temperatures. In \figdisp{fig:BHZHFSE}c, the DCA simulation on an $N=2\times2$ cluster benchmarks well with DQMC results on a $N=6\times6$ cluster at high temperature $\beta=4$. In addition, it shows no indication of a gap opening up to temperatures as low as $\beta=20$.

\clearpage

\subsection{DQMC simulations for the BHZ-HH model at different $U$}

\begin{figure}[ht!]
    \centering
    \includegraphics[width=0.8\textwidth]{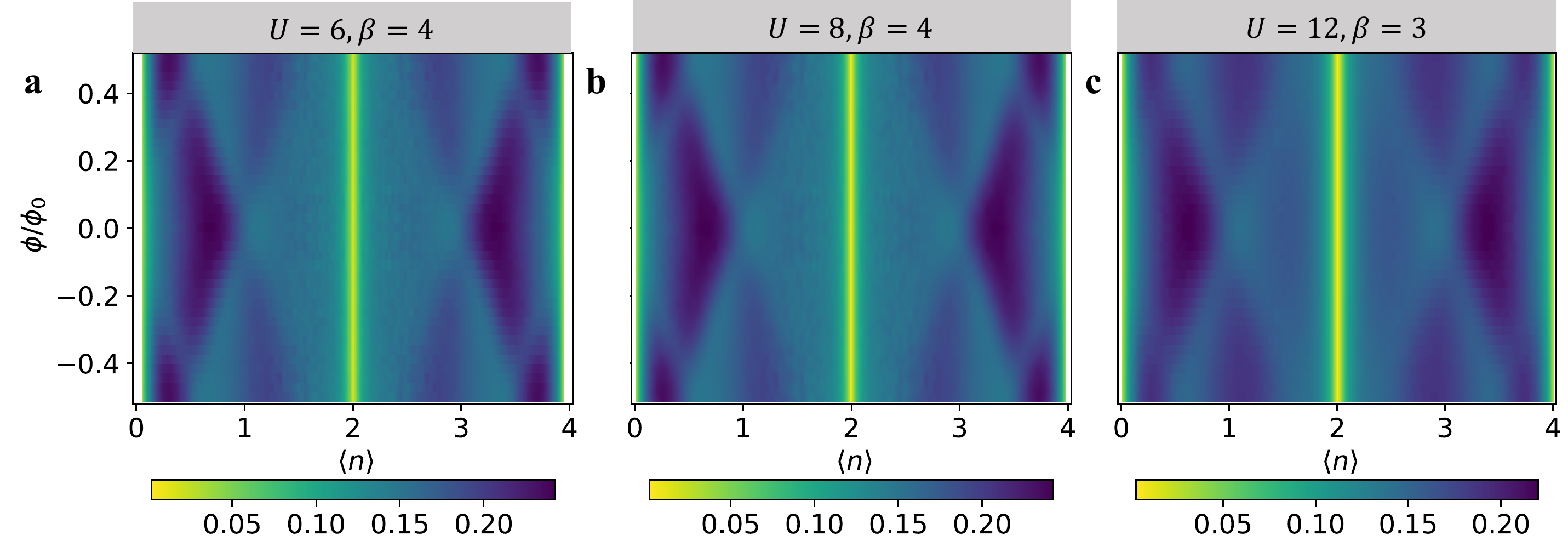}
    \caption{The compressibility as a function of density and magnetic flux for the BHZ-HH model ($M=1$) at {\bf a} $U=6,\beta=4$, {\bf b} $U=8,\beta=4$ and ({\bf c}) $U=12,\beta=3$.
    }
    \label{fig:BHZvaryU}
\end{figure}

\begin{figure}[ht!]
    \centering
    \includegraphics[width=0.4\textwidth]{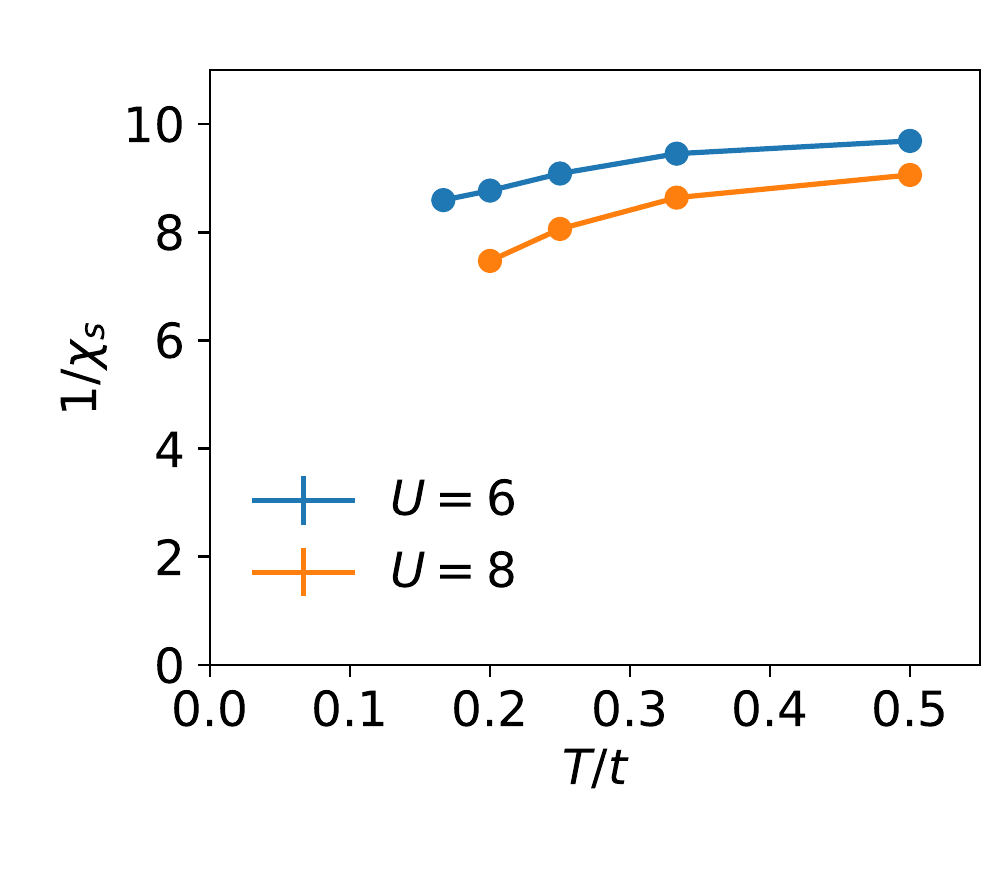}
    \caption{The inverse spin susceptibility as a function of temperature for the BHZ-HH model ($M=1$) at $U=6$ and $U=8$.
    }
    \label{fig:chisBHZU6U8}
\end{figure}

The previous subsection showed that the gap barely opens in the DCA simulations for the BHZ-HH model ($M=1$) at $U=8$. Here from the compressibility plots in \figdisp{fig:BHZvaryU}, we find that the non-trivial topology already emerges at quarter-filling for an interaction strength as small as $U=6$, thus justifying a semimetallic state with a high-temperature QSH feature. We also show the temperature-dependent inverse spin susceptibility at $U=6$ and $8$ in \figdisp{fig:chisBHZU6U8}. In either case, the spin susceptibility is unlikely to diverge at finite temperatures compared to the insulating case in Fig.~2a and Fig.~5d in the main text, supporting the semi-metallic state with no gap opening.

\clearpage

\subsection{Finite-size analysis on the BHZ-HH model}

\begin{figure}[ht!]
    \centering
    \includegraphics[width=0.9\textwidth]{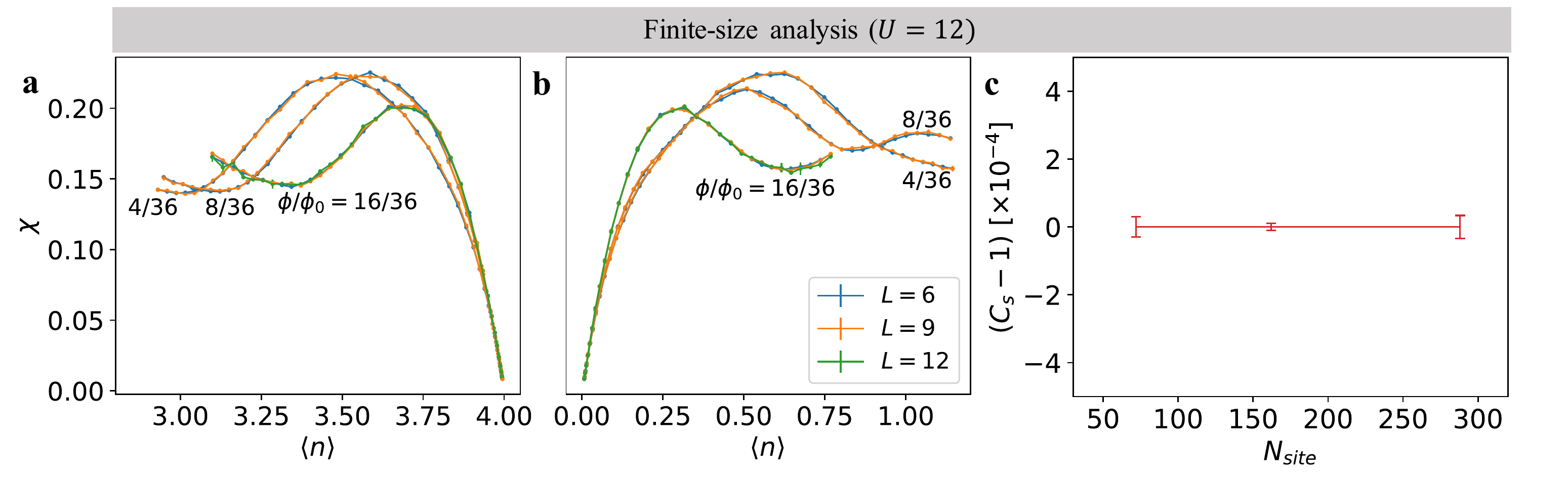}
    \caption{Panels {\bf a} and {\bf b} show the TRI compressibility and compressibility respectively at low hole and electron density for the BHZ-HH model, at varying cluster sizes $N_\text{site}=L\times L$ under different magnetic fluxes as labeled. They share the same legend. Panel {\bf c} presents the spin Chern number extracted for different cluster sizes. The temperature is $\beta=3/t$.
    }
    \label{fig:FSA}
\end{figure}

To corroborate our finding of QSH effects driven by Mottness at quarter-filling, we conducted a finite-size analysis on the TRI compressibility (\figdisp{fig:FSA}(a)) and the compressibiility (\figdisp{fig:FSA}(a)) of the BHZ-HH model (\figdisp{fig:FSA}(b)) at low hole and electron density respectively for a range of filling that covers the feature from $3/4$- and $1/4$-fillings with $U=12$. In \figdisp{fig:FSA}(c), we present the TRI compressibility at different magnetic fluxes $\phi/\phi_0=4/36,8/36,16/36$ and system sizes $N_\text{site}=L^2$ with $L=6,9,12$. The collapse of the curves regardless of system size (panels (a) and (b)) suggests that the dip feature at $1/4$- and $3/4$-filling is fundamental and survives the thermodynamic limit. As a consequence, it makes sense to extract a spin Chern number.  We find that for all the system sizes studied,  $C_s=1$ as depicted in \figdisp{fig:FSA}(c).

\subsection{Lower temperature results for BHZ-HH model}
\begin{figure}[ht]
    \centering
    \includegraphics[width=0.6\textwidth]{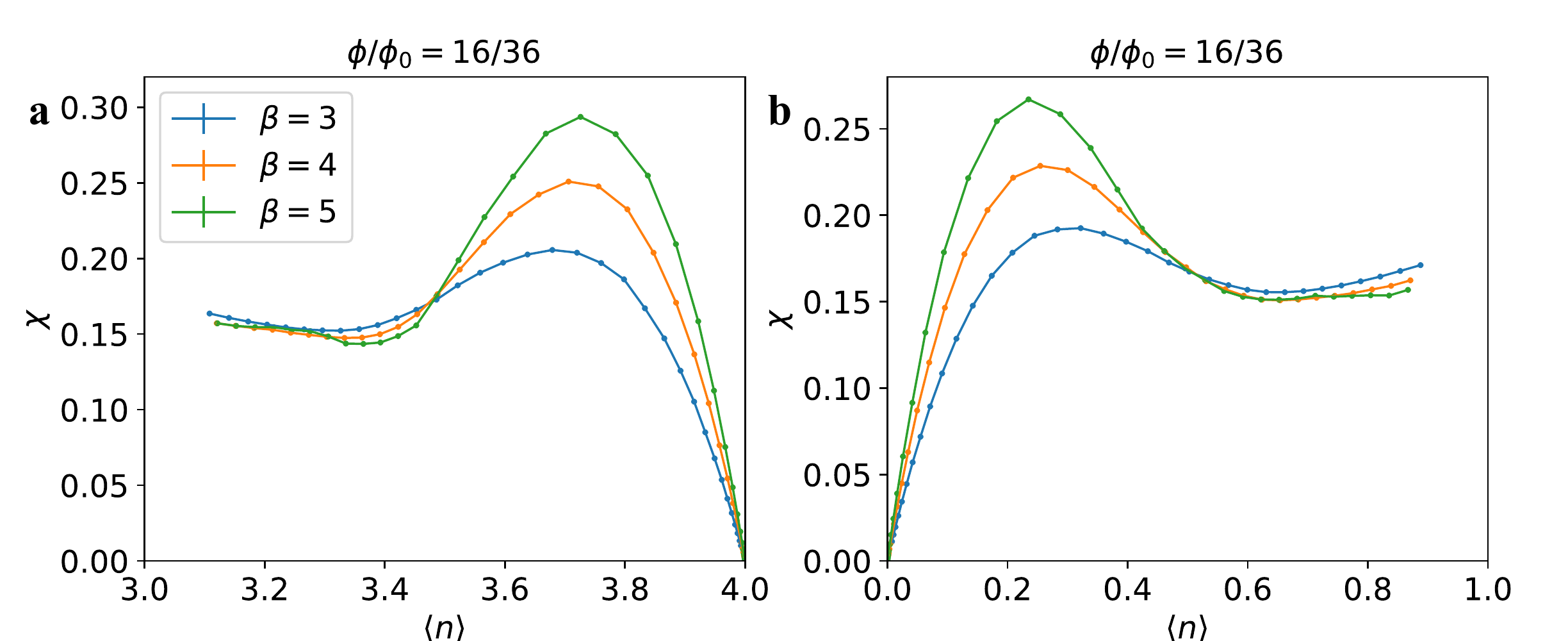}
    \caption{The TRI compressibility ({\bf a}) and compressibility 
 ({\bf b}) at low hole and electron density of the BHZ-HH model respectively, under a magnetic flux $\phi/\phi_0=16/36$ with different inverse temperatures $\beta=3,4,5$. The interaction strength is $U=8$.}
    \label{fig:lowerT}
\end{figure}

To explore lower-temperature physics, we are restricted to low density and a weaker interaction $U=8$. In \figdisp{fig:lowerT}, we show the TRI compressibility and compressibility at $\phi/\phi_0=16/36$ for the BHZ-HH model, though at low hole and electron density respectively, enough to capture the interested feature from quarter-fillings. In either case, the dips (representing the $C^{\text{TRI}}=1$ valley for the TRI compressibility in  \figdisp{fig:lowerT}(a) and the zero-mode Landau levels for the compressibility in panel \figdisp{fig:lowerT}(b)) don't decrease evidently despite the metallic peaks nearby rising up. This observation also supports a semimetallic state.

\clearpage

\subsection{Real-space spin correlation at half-filling under strong correlation}

\begin{figure}[ht!]
    \centering
    \includegraphics[width=0.5\textwidth]{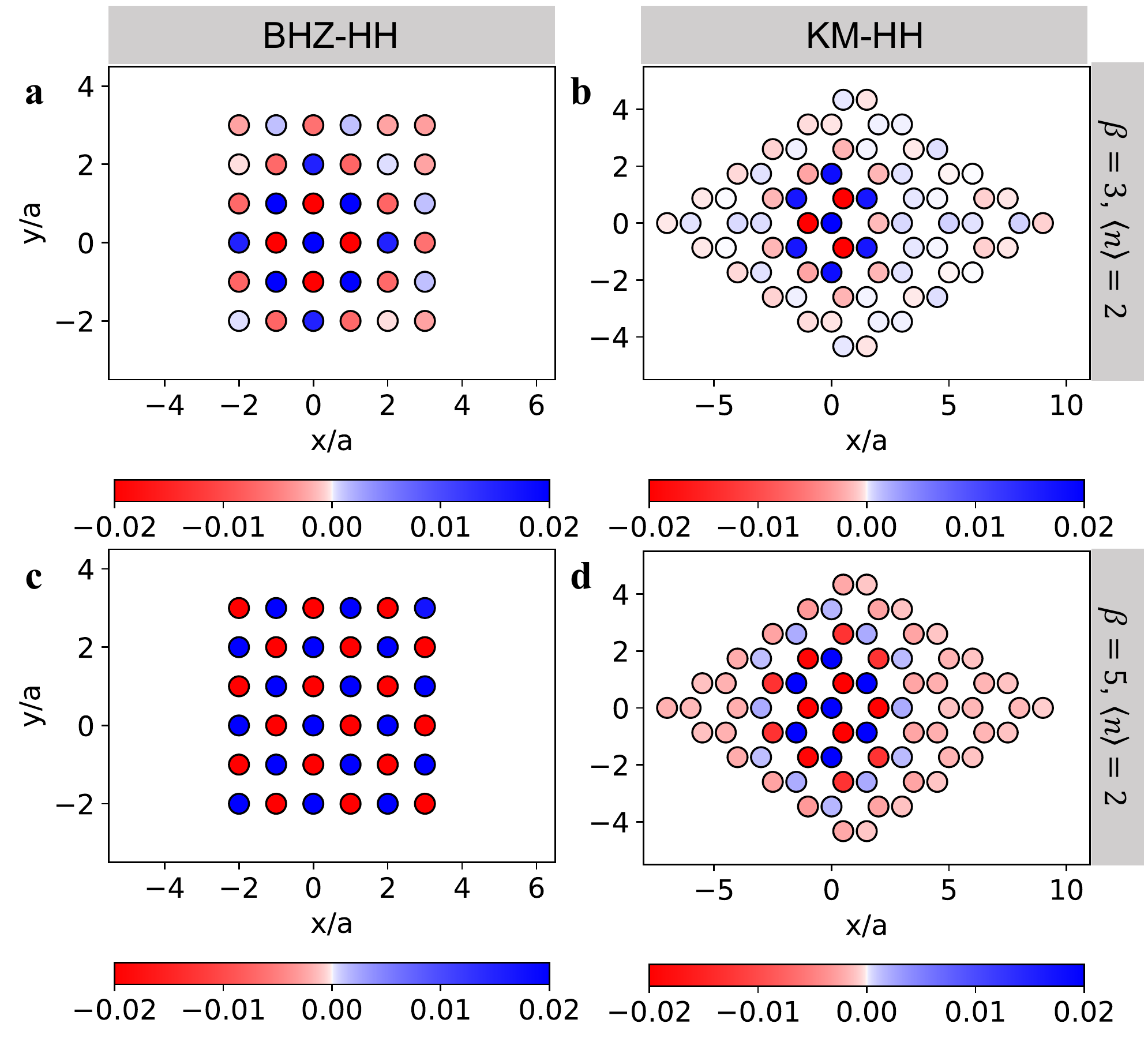}
    \caption{The static spin correlation at half-filling ($\langle n\rangle=2$) in real spaces respectively for BHZ-HH ({\bf a}, {\bf c}) at $M=1$ and KM-HH ({\bf b}, {\bf d}) models at $\psi=0.5,t^\prime=0.3$ both with $U=12$ (in the model-specific energy scale) and zero magnetic field. The inverse temperatures are $\beta=3$ for panels ({\bf a}, {\bf b}) and $\beta=5$ for panels ({\bf c}, {\bf d}).
    }
    \label{fig:sup_realspace}
\end{figure}

In the presence of strong correlation, at half-filling the non-interacting QSH order is destroyed and turns into a topologically trivial Mott insulator. We expect the Mott insulator to show anti-ferromagnetism (AF) robust to an external magnetic field in a bipartite lattice. However, Fig.~3(c) in the main text shows a peak in the spin correlation (${\bf Q}=0$) at half-filling, which is inconsistent with this expectation. To further explore this issue, we look into its zero-frequency spin correlation in real and momentum space at two temperatures $\beta=3$ and $\beta=5$, shown in \figdisp{fig:sup_realspace}. Even at the relatively higher temperature of $\beta=3$, the system already shows an AF pattern in the central region. As the temperature decreases, the AF region enlarges and further dominates the cluster. Thus, eventually at low enough temperature, AF order prevails in the Mott insulator, as expected. Note that here the smallest magnetic flux $\phi/\phi_0=1/36$ is used to reduce the finite-size effect at lower temperatures \cite{mfp}. This does not affect our conclusion because the influence from the magnetic field is negligible compared to the AF order. 

\clearpage

\section{Calculation of spin susceptibility in the presence of Hatsugai-Kohmoto interaction}

The BHZ-Hatsugai-Kohmoto (HK) model is
\beq
\begin{aligned}
H&=\sum_{{\bf k},\sigma}\big[(\varepsilon_{+,{\bf k},\sigma}-\mu)n_{+,{\bf k},\sigma}+(\varepsilon_{-,{\bf k},\sigma}-\mu)n_{-,{\bf k},\sigma}\big]+
U\sum_{\bf k}(n_{+,{\bf k},\uparrow}n_{+,{\bf k},\downarrow}+n_{-,{\bf k},\uparrow}n_{-,{\bf k},\downarrow}). \label{HHK}
\end{aligned}
\eeq

We are mostly interested in zero temperature, strong correlations and low filling up to $1/4$. So then the singly and doubly occupied topological upper band is completely unoccupied. We can drop that for the current calculation. Adding a small Zeeman field, the low energy effective Hamiltonain becomes

\beq
\begin{aligned}
H&=\sum_{{\bf k},\sigma}\big[(\varepsilon_{-,{\bf k},\sigma}-\mu)n_{-,{\bf k},\sigma}\big]+
U\sum_{\bf k}n_{-,{\bf k},\uparrow}n_{-,{\bf k},\downarrow}-h\sum_{\bf k}(n_{-,{\bf k},\uparrow}-n_{-,{\bf k},\downarrow}). \label{HHK2}
\end{aligned}
\eeq
Note that $h\sum_{\bf k}(n_{-,{\bf k},\uparrow}-n_{-,{\bf k},\downarrow})=h\sum_{\bf i}(n_{-,{\bf i},\uparrow}-n_{-,{\bf i},\downarrow})$. In Eq.~(\ref{HHK2}), ${\bf k}$ is still a good quantum number. Then we can write down the partition function:
\beq
Z=\prod_{\bf k} Z_{\bf k},~~ Z_{\bf k}=1+\exp[-\beta(\varepsilon_{-,{\bf k}}-\mu-h)]+\exp[-\beta(\varepsilon_{-,{\bf k}}-\mu+h)]+\exp[-\beta(2\varepsilon_{-,{\bf k}}-2\mu+U)]
\eeq
in which $\varepsilon_{-,{\bf k}}=\varepsilon_{-,{\bf k},\sigma}$ since the dispersion is independent of spin. Then the average occupation at each $k$ and $\sigma$ is
\beq
\langle n_{{\bf k},\sigma} \rangle=\frac{1}{Z_k} (\exp[-\beta(\varepsilon_{-,{\bf k}}-\mu-\sigma h)]+\exp[-\beta(2\varepsilon_{-,{\bf k}}-2\mu+U)])
\eeq
The magnetic moment is 
\beq
\langle m \rangle=\frac{1}{2}(\langle n_\uparrow \rangle-\langle n_\downarrow \rangle)
=\frac{1}{2}\sum_{\bf k}(\langle n_{{\bf k},\uparrow} \rangle-\langle n_{{\bf k},\downarrow} \rangle)
=\frac{1}{2}\sum_{\bf k}\frac{1}{Z_k}(\exp[-\beta(\varepsilon_{-,{\bf k}}-\mu-h)]-\exp[-\beta(\varepsilon_{-,{\bf k}}-\mu+ h)]).
\eeq
It is obvious that $\langle m \rangle\rightarrow 0$ as $h\rightarrow 0$, as expected. Next we calculate the magnetic susceptibility 
\beq
\chi_s=\frac{d\langle m \rangle}{dh}\big|_{h\rightarrow0}=\sum_{\bf k}\frac{\beta \exp[-\beta(\varepsilon_{-,{\bf k}}-\mu)]}{Z_k(h=0)}.
\eeq
From the main text, $\varepsilon_{-,{\bf k}}<0$ for all ${\bf k}$ and $\mu=0$ at $1/4$-filling. Then the magnetic susceptibility  becomes
\beq
\chi_s(\langle n \rangle=1)
=\sum_{\bf k}\frac{\beta \exp[-\beta \varepsilon_{-,{\bf k}}]}{1+2\exp[-\beta \varepsilon_{-,{\bf k}}]+\exp[-\beta (2\varepsilon_{-,{\bf k}}+U)]}
=\sum_{\bf k}\frac{\beta \exp[-\beta \varepsilon_{-,{\bf k}}]}{1+2\exp[-\beta \varepsilon_{-,{\bf k}}]}.
\eeq
where at low temperature, the term $\exp[-\beta (2\varepsilon_{-,{\bf k}}+U)]$ is dropped for large $U$. At the zero temperature limit ($\beta\rightarrow\infty$),
\beq
\chi_s(\langle n \rangle=1,\beta\rightarrow\infty)=\frac{A_B \beta}{2} \label{SSHK}
\eeq
where $A_B$ is the area of the Brillouin zone. This result can be generalized: for $\langle n \rangle\leq2$, $\chi_s(\beta\rightarrow\infty)\sim \beta$. To compare, the susceptibility for the tight-binding model is 
\beq
\chi_s(\mu, U=0)=\sum_{\bf k}\frac{\beta \exp[-\beta (\varepsilon_{-,{\bf k}}-\mu)]}{1+2\exp[-\beta (\varepsilon_{-,{\bf k}}-\mu)]+\exp[-2\beta (\varepsilon_{-,{\bf k}}-\mu)]}.
\eeq
Setting $\mu=0$ (half-filling $\langle n \rangle=2$) and taking the zero temperature limit, it becomes
\beq
\chi_s(\langle n \rangle=2, U=0, \beta\rightarrow\infty)=\sum_{\bf k}\frac{\beta \exp[-\beta \varepsilon_{-,{\bf k}}]}{\exp[-2\beta\varepsilon_{-,{\bf k}}]}\big|_{\beta\rightarrow\infty}\rightarrow0.
\eeq
The divergence of the spin susceptibility in \disp{SSHK} indicates the instability towards a spin polarized ferromagnetic state, driven by strong correlation.
\clearpage

\section{Bilayer KM model}

\begin{figure}[ht]
    \centering
    \includegraphics[width=0.8\textwidth]{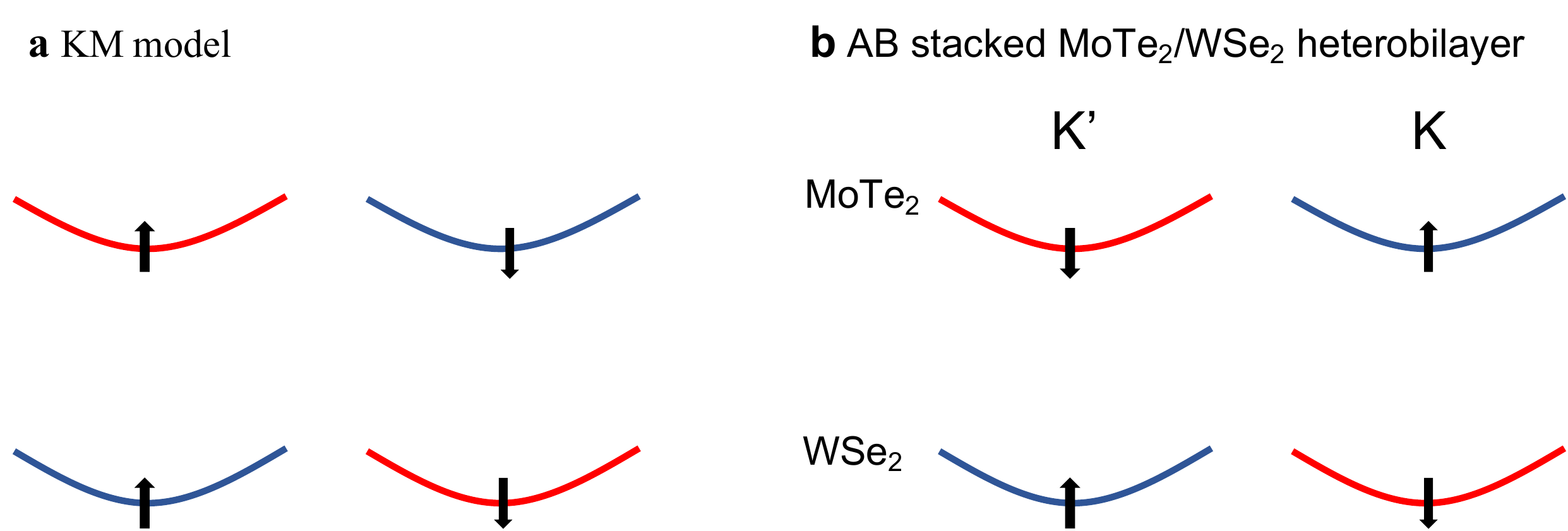}
    \caption{The schematic band structures of {\bf a} KM model and {\bf b} AB-stacked MoTe$_2$/WSe$_2$ heterobilayer. Blue and red colors represent Chen number $C=1$ and $-1$ respectively. $K$ and $K'$ are valley degrees of freedom.}
    \label{fig:sup_KMvsAB}
\end{figure}

\begin{figure}[ht]
    \centering
    \includegraphics[width=0.8\textwidth]{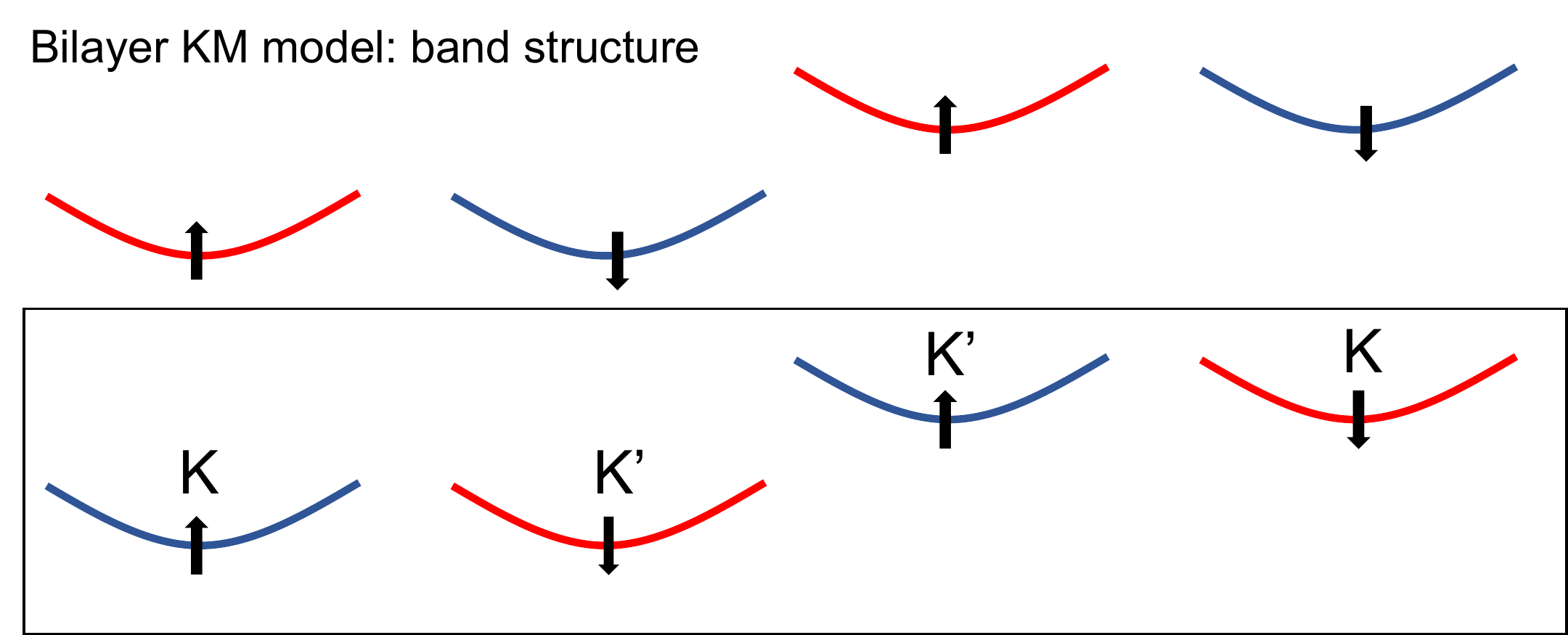}
    \caption{The schematic band structure of the bilayer KM model. Blue and red colors represent Chen number $C=1$ and $-1$ respectively.} 
    \label{fig:sup_biKM}
\end{figure}

\begin{figure}[ht]
    \centering
    \includegraphics[width=0.8\textwidth]{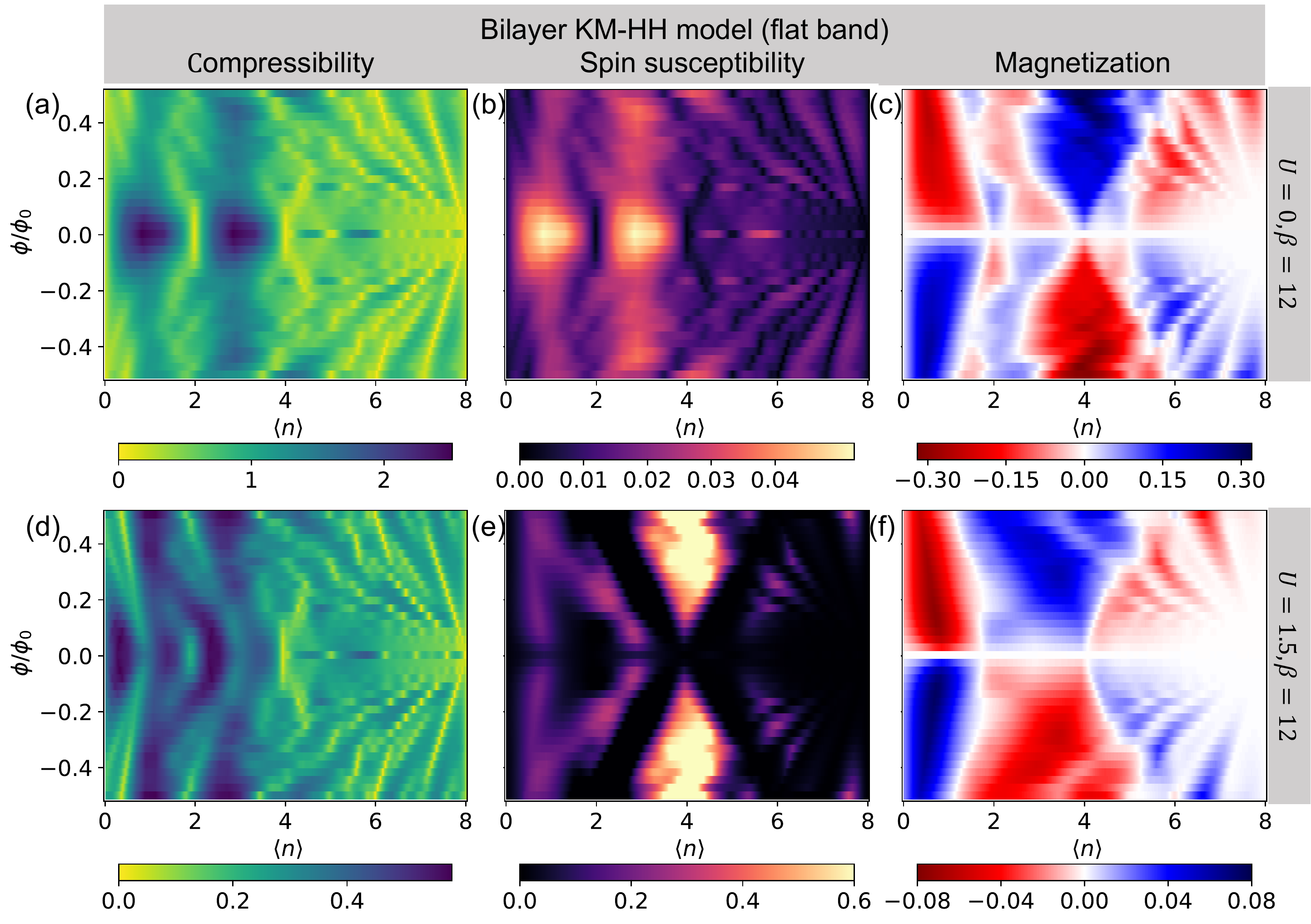}
    \caption{DQMC results for the flat-band bilayer KM-HH models at $U = 0$ (first row) and $U/t = 1.5$ (second row). Each row shows the compressibility, spin susceptibility and magnetization, all as a function of density and magnetic flux. The parameters are $t=1$, $t'/t= 0.3$, $\psi=2.54$, $t_\perp/t=0.3$, $V/t=0.4$ and $\beta=12/t$.} 
    \label{fig:sup_biKM2}
\end{figure}

In \figdisp{fig:sup_KMvsAB}, comparing the schematic band structure of the KM model with that in the AB-stacked MoTe$_2$/WSe$_2$ heterobilayer\cite{Taozui} (blue and red colors are for Chen number $C=1$ and $-1$ respectively), we find that the only difference is that they have opposite spin for the top Chern bands. This contrast leads to qualitatively different physics. Filling all four KM bands in \figdisp{fig:sup_KMvsAB}(a) gives rise to a trivial band insulator, while filling all four moir{\'e} bands in \figdisp{fig:sup_KMvsAB}(b) results in a double QSH insulator with a spin Chern number $C_s=4$ since $C_{\uparrow}=2$ and $C_{\downarrow}=-2$. The latter is inherently contradictory as non-trivial topology cannot result in the band insulator limit.  Consequently, it is not possible to construct a four-band tight-binding model to describe this physics in AB-stacked heterobilayer. The principle for constructing a tight-binding model is that the system becomes a trivial band insulator if all bands are filled. Therefore, to describe the physics in the AB-stacked MoTe$_2$/WSe$_2$ heterobilayer, we use a bilayer KM model (with eight bands) whose Hamiltonian is
\beq
H=H_{\text{KM}_1}+H_{\text{KM}_2}+t_\perp \sum_{i,\sigma}(c_{1i\sigma}^\dagger c_{2i\sigma}+c_{2i\sigma}^\dagger c_{1i\sigma})+V\sum_{i,\sigma}(c_{1i\sigma}^\dagger c_{1i\sigma}-c_{2i\sigma}^\dagger c_{2i\sigma}),
\eeq
where $i$ is the site label for each layer, $\sigma$ represents the spin and the numbers $1$ and $2$ are layer indices, $V$ is the voltage difference between layers. $H_{\text{KM}_1}$ is the same as $H_{\text{KM}_2}$ with only different layer indices. In the flat-band limit, the inter-layer hopping $t_\perp$ lifts the degeneracy of the bottom (or top) orbitals between two KM layers while the Chern number for each band with a specific spin is unchanged, as shown in \figdisp{fig:sup_biKM}(a). Then the lower four bands are equivalent to the four moir{\'e} bands in the AB-stacked MoTe$_2$/WSe$_2$ heterobilayer (\figdisp{fig:sup_KMvsAB}) for which the valley labels can be assigned. Without interaction, this model shows a QSH effect at quarter-filling ($\langle n \rangle=2$) and a double QSH effect at half-filling  ($\langle n \rangle=4$) in \figdisp{fig:sup_biKM2}(a-c) with an example flat-band parameter set $t'/t= 0.3$, $\psi=2.54$, $t_\perp/t=0.3$, $V/t=0.4$ and $\beta=12/t$.

We conduct the DQMC simulation for the corresponding bilayer flat-band KM-HH model at an intermediate Hubbard interaction $U=1.5t, \beta=12/t$ and present the compressibility, spin susceptibility and magnetization in \figdisp{fig:sup_biKM2}(d-f) respectively. The system exhibits a QAH effect with spin polarization at $1/8$-filling ($\langle n \rangle=1$) and a QSH effect at quarter-filling ($\langle n \rangle=1$), as expected given the similarity between single-layer and bilayer KM-HH models. Since the Hubbard interaction mixes the non-interacting bands and the resulting QAH state is spin polarized, this necessarily yields valley-coherence given the valley assignment in \figdisp{fig:sup_biKM}(a).

The presence of an emergent topologically non-trivial state at $3/8$-filling ($\langle n \rangle=3$) is special to the bilayer model. It has the feature of a QAH effect, namely the single Landau level in the compressibility (\figdisp{fig:sup_biKM2}(d)) accompanied by the peak in the spin susceptibility (\figdisp{fig:sup_biKM2}(e)). It is also likely to have the QSH feature. This obtains because the system can not be fully polarized at $\langle n \rangle=3$. Some band must be doubly occupied, thereby explaining the white region in the magnetization in (\figdisp{fig:sup_biKM2}(f)). Also, the helical currents from the $\langle n \rangle=2$ QSH state probably play a role in the $\langle n \rangle=3$ edge state because their edge dispersion extends to the upper bands, which would not be filled until $\langle n \rangle>4$. Therefore, for the state at $\langle n \rangle=3$, we have either $C_\uparrow=2$, $C_\downarrow=-1$ or $C_\uparrow=1$, $C_\downarrow=-2$ depending on the polarization while the $\langle n \rangle=1$ state has either $C_\uparrow=1$, $C_\downarrow=0$ or $C_\uparrow=0$, $C_\downarrow=-1$. 

\bibliography{reference}